\documentclass[preprint2,showpacs,showkeys]{revtex4-1}
\usepackage{graphicx}
\usepackage{dcolumn}
\usepackage{bm}%
\usepackage{float}
\usepackage[caption = false]{subfig}

\begin{document}

\title{Effect of magnetic field on the burning of a neutron star}

\date{\today}

\author{Ritam Mallick} 
\email{mallick@iiserb.ac.in}
\affiliation{Indian Institute of Science Education and Research Bhopal, Bhopal, India}
\author{Amit Singh}
\affiliation{Indian Institute of Science Education and Research Bhopal, Bhopal, India}

\begin{abstract}
In this article, we present the effect of a strong magnetic field in the burning of a neutron star (NS). We have used relativistic magneto-hydrostatic (MHS) conservation equations for studying the PT from nuclear matter (NM) to quark matter (QM). 
We found that the shock-induced phase transition (PT) is likely if the density of the star core is more than three times nuclear saturation ($\rho_s$) density. 
The conversion process from NS to quark star (QS) is found to be an exothermic process beyond such densities. 
The burning process at the star center most likely starts as a deflagration process. However, there can be a small window at lower densities where the process can be a detonation one.  
At small enough infalling matter velocities the resultant magnetic field of the QS is lower than that of the NS. However, for a higher value of infalling matter
velocities, the magnetic field of QM becomes larger. Therefore, depending on the initial density fluctuation and on whether the PT is a violent one or not the QS 
could be more magnetic or less magnetic. 
The PT also have a considerable effect on the tilt of the magnetic axis of the star. For smaller velocities and densities the magnetic angle are not affected much but for higher
infalling velocities tilt of the magnetic axis changes suddenly. The magnetic field strength and the change in the tilt axis can have a significant effect on the observational aspect
of the magnetars.
\end{abstract}



%

\maketitle

\section{Introduction}
\label{intro} 
One of the most challenging aspects of astrophysics is the study and understanding of compact objects. 
Compact objects usually refer to the family of white dwarfs, compact stars (CSs) and black hole, usually formed after the gravitational collapse of a dead star. Among compact 
objects, CSs 
(otherwise commonly known as neutron stars (NS) or quark stars (QS)) bear a
special significance in astrophysics since in addition to their own
importance they also serve as a tool to improve the understanding of
nuclear matter (NM) and possibly quark matter (QM) at enormous densities and low temperatures (see, e.g., \cite{weber,glen}).
Thus the compact star serves as an ideal complementary approach to the study of high-temperature relativistic heavy-ion collisions. 

Our understanding of compact stars has changed in the last fifty years, beginning with the discovery of pulsars \cite{hewish} whilst connecting them with NSs \cite{gold}. 
It was well understood that pulsars are nothing but spinning CSs mostly emitting x-rays and radio waves. The central density of CS is inferred to 
be as high as $3-10$ times the nuclear saturation density.
Over time, different equation of state (EoS) of matter at such high density have been proposed and are being continuously refined.
One of the most exciting aspects arising from such high-density stars is the occurrence of QM in their cores where confinement to deconfinement transition takes place, resulting in QS. Therefore, CS can be of two types \\
a) NS composed entirely of NM \\
b) QS which have some quantity of deconfined QM in them. \\

While the nuclear and quark models have improved over the years, significant advancements appeared from the astrophysical observations.
The change has been more rapid in the last decade when the discovery and timing observation of pulsars gained acceleration due to advent of the new generation of 
space-based X-ray and gamma-ray satellites (Einstein / EXOSAT). Important observations also came from the ROSAT observatory. However, a new era of thermal radiation 
observation started after the launch of CHANDRA and XMM-Newton Observatory. With
improved telescopes and interferometric techniques, the number of observed binary pulsars are
continuously increasing. To date, we know precise masses of about $35$ pulsars spanning the range from 
$1.15 M_{\odot}$to $2.01 M_{\odot}$. The radius measurement is not as precise as the masses. However, it is widely accepted that they must lie in the range between $9-13$ km. 
The knowledge of heaviest NS, PSR J1614-2230 and PSR J0348+0432 \cite{demorest,antonidis} and connecting them with the existing
radius bound already places a significant constraint on the EoS of matter at these extreme densities.

The possible existence of both NS and QS has been proposed long back \cite{itoh,bodmer,witten}. The conversion of an NS to a QS 
is likely through a deconfined phase transition (PT). {\bf The PT can occur either soon after the formation of the NS in a
supernovae explosion or during the later time through a first order PT or a smooth crossover transition.} The phase transformation is usually assumed to begin at the center of a 
star when the density increases beyond the critical density. Several processes can trigger PT: slowing down of the rotating star \cite{glen1}, accretion of matter on the stellar surface \cite{alcock}.  
The cooling of a neutron star by magnetic field decay \cite{geppert} can also trigger this process. Such a PT is characterized by a significant energy release in the form of
latent heat, which is accompanied by a neutrino burst, thereby cooling the star. Corresponding star transformations
should lead to interesting observable signatures like $\gamma$ -ray bursts \cite{drago,bombaci1,berezhiani,mallick-sahu}, changes in the cooling rate \cite{sedrakian},
and the gravitational wave (GW) emission \cite{abdikamalov}.

The dynamical study of PT is somewhat uncertain and even controversial \cite{horvath1}. In literature one can find two very
different scenarios: (i) the PT is a slow deflagration process and never a detonation \cite{drago1} and (ii) the PT from confined to
deconfined matter is a fast detonation-like process, which lasts about 1 ms \cite{bhat1}. If the process is quick burning and very violent
(detonation) one, there can be robust GW signals arising from them which can be detected at least in the second or
third generation of VIRGO and LIGO GW detectors \cite{abdikamalov,lin}. The earliest calculations \cite{olinto} assumed the conversion to proceed via slow combustion, 
where the conversion process depends strongly on the temperature of the star. However, Horvath \& Benvenuto \cite{horvath} later studied the stability of this conversion process, 
and found that under the influence of gravity the conversion process becomes unstable and the slow combustion can become a fast detonation.
A relativistic calculation was performed \cite{cho} to determine the nature of the conversion process, employing the energy-momentum and baryon number conservation
(also known as the Rankine-Hugoniot condition). A recent calculation of the burning process for violent shocks has also been studied \cite{igor}.
However, there is still no consensus about the nature of the conversion process.

Another unique feature of compact stars is the presence of ultra-strong magnetic field at their surface.
The surface field strength for almost all pulsars are of the order of $10^8 - 10^{12}$ G. However, recent observations of several new pulsars, 
namely some anomalous X-ray pulsars (AXP) and soft-gamma repeaters (SGR), have been identified to have much stronger
surface magnetic fields \cite{kulkarni,murakami}. Such pulsars with strong magnetic
fields are separately termed as magnetars \cite{duncan,thompson}. Such a field is usually estimated from observation of the NS period and their derivative.
It has also been attributed that the observed giant flares, SGR 0526-66, SGR 1900+14 and SGR 1806-20, are the manifestation
of such strong surface magnetic field in those stars. While magnetic fields as high as $10^{15}$ G have been inferred at the surface of magnetars \cite{duncan,paczynski,melatos},
there is indirect evidence for fields as high as $10^{16}$ G inside the star \cite{makishima}. It is believed that at the dense cores of such stars the magnetic field
is a few order higher and in theoretical calculation one often assumes the magnetic field to be of the order of $10^{18}$ G \cite{mallick-deform,dexheimer}.

The origin of such a high magnetic field is still unknown. The magnetic field of regular old pulsars is attributed to the conservation of the magnetic flux during 
core collapse of the supernovae. However, they are unable to explain the strong surface fields of magnetars.
The idea by Thompson and Duncan \cite{duncan}, suggest a dynamo process by combining convection and differential rotation in hot proto-neutron
stars which can build up a field of strength of $10^{15}$ G. Recently it was proposed that magneto-rotational instability (MRI) and MRI driven dynamo in hot proto-neutron 
stars can amplify average magnetic field strength to very high values in quite short time \cite{akiyama,obergaulinger,sawai,mosta,rembiasz}. 
Whatever may be the origin of such 
magnetic fields it is clear that they will have a significant impact on the physical aspect of such stars.

This present work aims to study the effect of such a strong magnetic field in the conversion of NS to QS. Instead of using the relativistic conservation condition we would 
employ magneto hydrostatic conservation condition in the Hoffmann -Teller (HT) frame \cite{mallick-tl}. We will treat the matter as an ideal fluid with infinite conductivity. We will mostly 
concern ourselves with the space-like shocks, where the shock propagates with a velocity less than the speed of light. 

In our investigation, we will assume that a PT takes place inside a cold NS. {\bf The PT is presumably a first order PT.} 
We assume that the formation of the new phase takes place at the center of the star due to a 
sudden fluctuation of the star density. The star then burns from the core to the periphery. The conversion process will be determined by the conservation equations and the EoS of the matter on either side of the front. 

The paper is organized as followed. In section 2 we discuss the effect of magnetic field on the EoS of a star, both before and after the PT keeping in mind the recent observational 
bound. The original star is of hadronic
matter whereas the final burned state is of deconfined QM. In section 3 we discuss the effect of magnetic field on the star structure. Next, in section 4 we present the 
MHS conservation equation for the space-like and time-like conversion.
In section 5 we show our results aiming to clarify and classify the conversion process. Finally, in section 6 we summarize our findings and discuss their potential 
astrophysical implications.

\section{Magnetic field induced EoSs}
\label{sec:1}

\begin{figure}
\subfloat[]{\includegraphics[width = 2.4in]{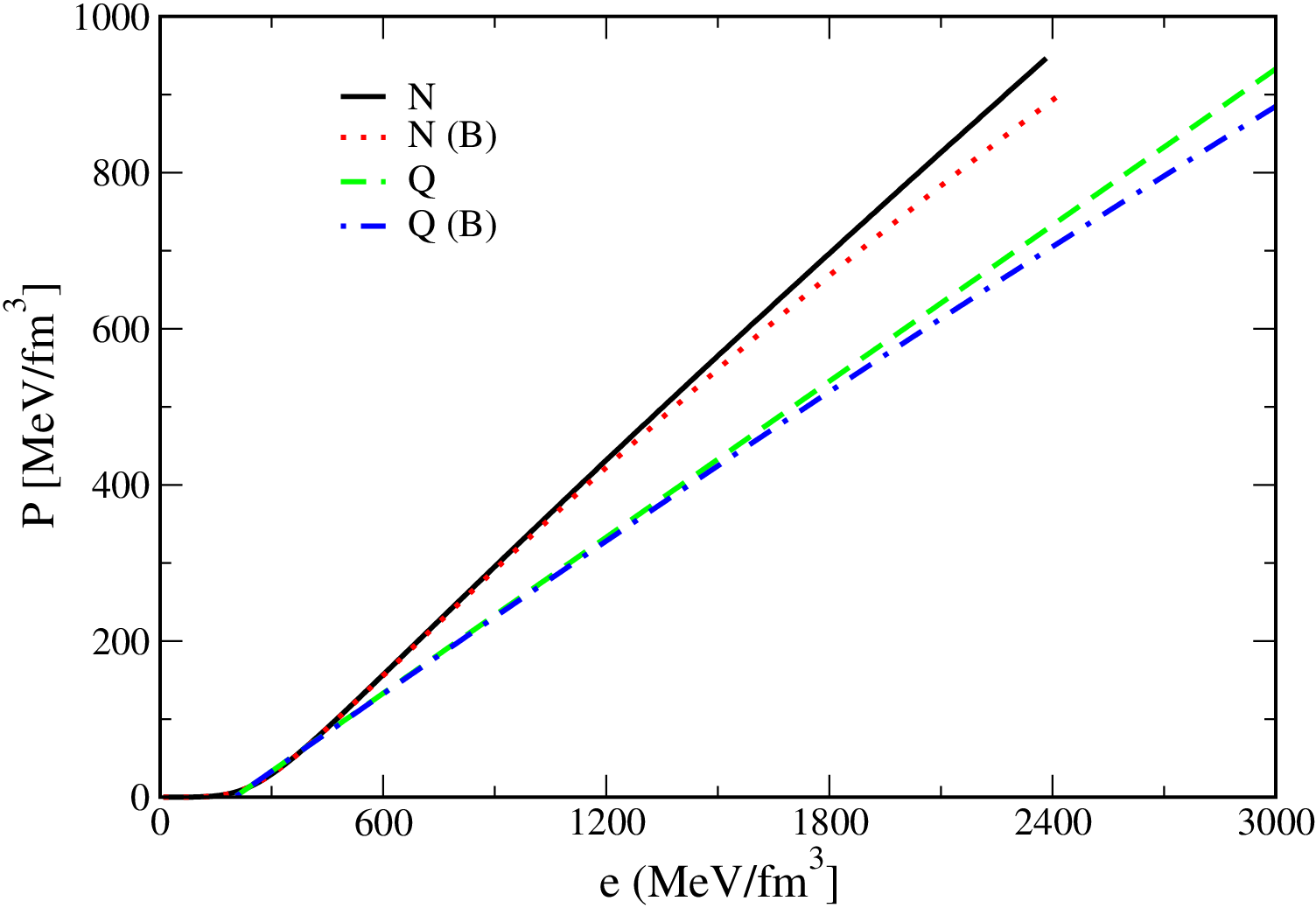}} \quad 
\subfloat[]{\includegraphics[width = 2.4in]{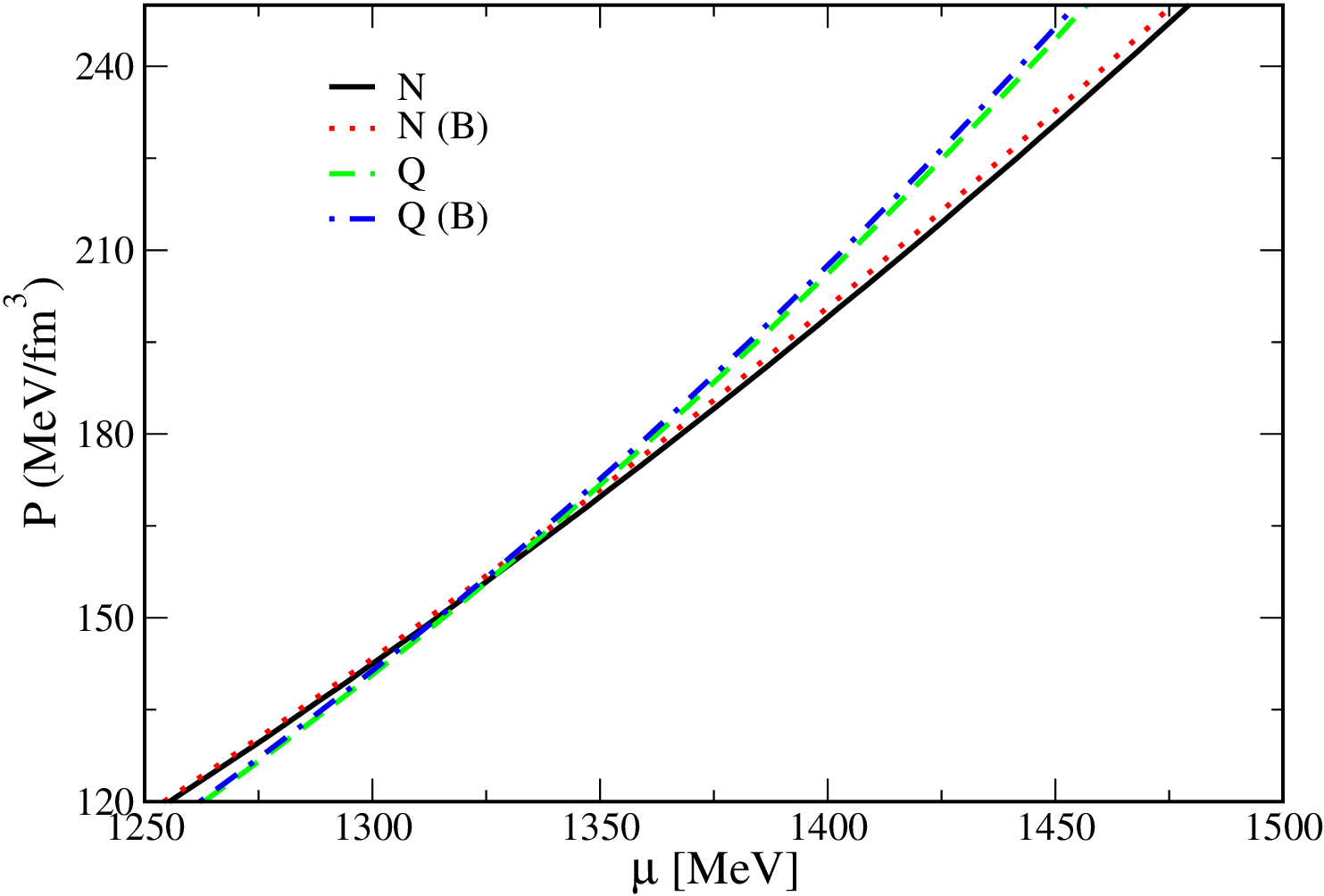}}
\caption{(Color online) a) EoS for NM and QM are shown. N stands for NM and Q for QM. For comparison curve with magnetic field induced EoS are also shown with B.
b) Pressures of NM and QM as functions of baryon chemical potential.  
The intersection point corresponds to the equilibrium PT from NM to QM. Similar curve with a magnetic field induced EoS are also plotted following similar prescription as in plot 1a.}
\label{eos}
\end{figure}

\begin{figure}
\centering
\includegraphics[width = 3.4in]{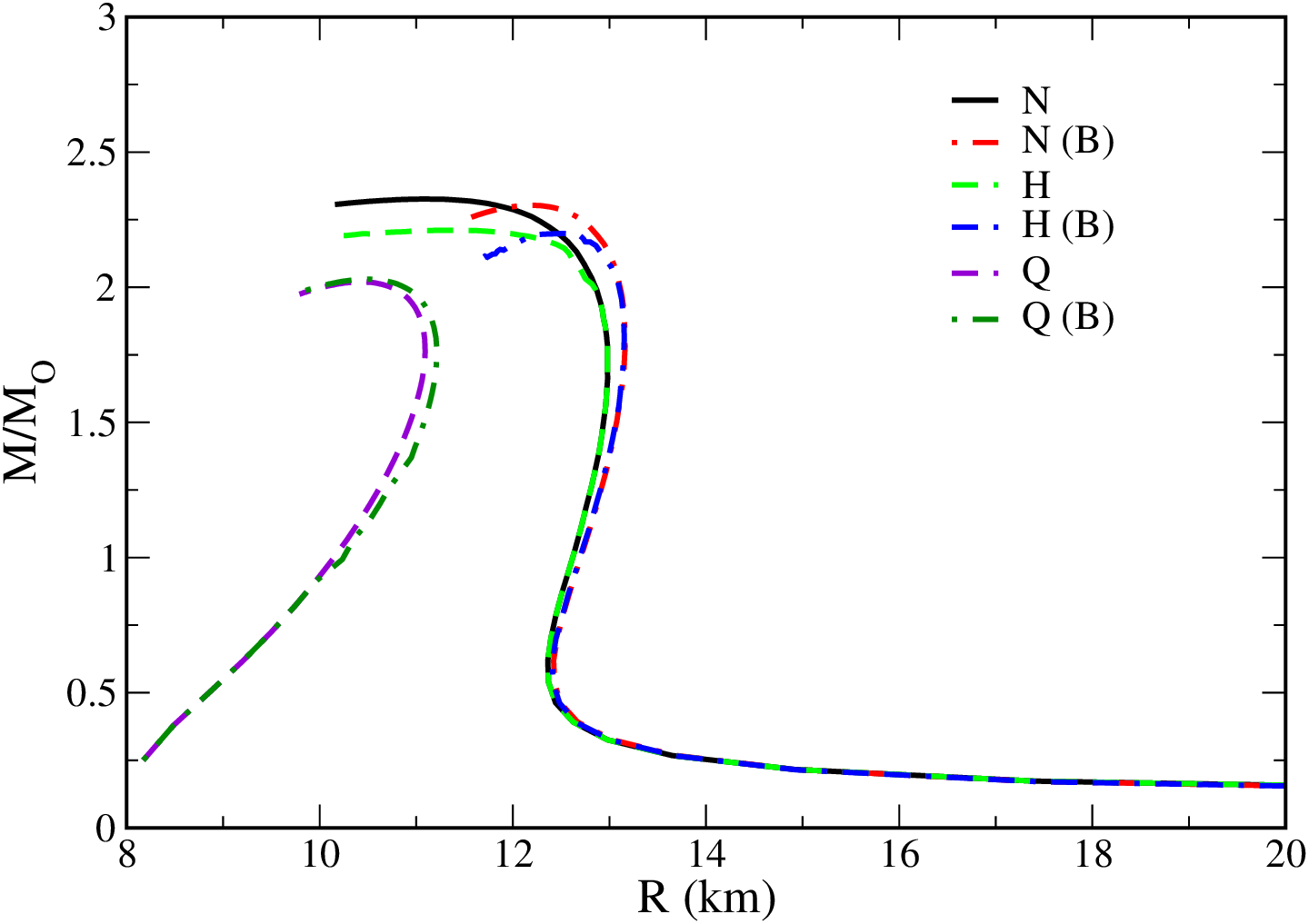}
\caption{(Color online) The mass-radius sequence of different compact stars (NS, HS and pure QS) obtained by solving the TOV equations (N stands for NS, H for HS and Q for QS). 
Also shown in the graphs are the mass-radius sequence of magnetized stars (symbolized with (B)).}
\label{mr}
\end{figure}

The PT is brought about by a sudden density fluctuation at the star core. This initiates a finite density and pressure fluctuation which propagates outwards.
This fluctuation is assumed to propagate along a single very thin layer, known as PT front, converting NM to QM. Therefore, to describe the properties of NM and QM, 
we need their corresponding EoSs. 
We employ such EoS which satisfies the current bound on the recent pulsar mass measurement.
We use zero temperature EoS as we assume that the PT takes place due to density fluctuation in any ordinary cold pulsar. However, the final burnt QM 
can have finite temperature depending on the EoS of matter on either side of the PT front.
For the hadronic phase, we adopt a relativistic mean-field approach which is generally used to describe
the NM in CS. The corresponding Lagrangian is given in the following form \cite{serot86,boguta,glendenning}($\hbar$=c=1)

 \begin{eqnarray} 
 {\cal L}_H = \sum_{n} \bar{\psi}_{n}\big[\gamma_{\mu}(i\partial^{\,\mu}  - g_{\omega n}\omega^{\,\mu} - 
\frac{1}{2} g_{\rho n}\vec \tau . \vec \rho^{\,\mu})- \\ \nonumber
\left( m_{n} - g_{\sigma n}\sigma \right)\big]\psi_{n} 
 + \frac{1}{2}({\partial_{\,\mu} \sigma \partial^{\,\mu} \sigma - m_{\sigma}^2 \sigma^2 } )- \frac{1}{3}b\sigma^{3}- \\ \nonumber
 \frac{1}{4}c\sigma^{4} - \frac{1}{4} \omega_{\mu \nu}\omega^{\,\mu \nu}+ 
\frac{1}{2} m_{\omega}^2 \omega_\mu \omega^{\,\mu} -\frac{1}{4} \vec \rho_{\mu \nu}.\vec \rho^{\,\mu \nu} + \\ \nonumber
\frac{1}{2} m_\rho^2 \vec \rho_{\mu}. \vec \rho^{\,\mu} 
+ \sum_{l} \bar{\psi}_{l}    [ i \gamma_{\mu}  \partial^{\,\mu}  - m_{l} ]\psi_{l}. 
\label{baryon-lag} 
\end{eqnarray}

The EoS contains only nucleons ($n$) and leptons ($l=e^{\pm},\mu^{\pm}$).
The leptons are assumed to be non-interacting, but the nucleons interact with the scalar $\sigma$ mesons, the isoscalar-vector $\omega_\mu$ mesons and the isovector-vector 
$\rho_\mu$ mesons. The fundamental properties of NM and that of finite nuclei are used to fit the adjustable parameter of the model.
In our present calculation, we use PLZ parameter set \cite{plz1,plz2}, which usually generates massive NSs, with $12.9$ km radius for $1.4 M_{\odot}$ star, which are in 
agreement with recent constraints of mass and 
radius \cite{steiner,lattimer}.

\begin{figure}
\centering
\vskip 0.2in
\includegraphics[width = 3.4in]{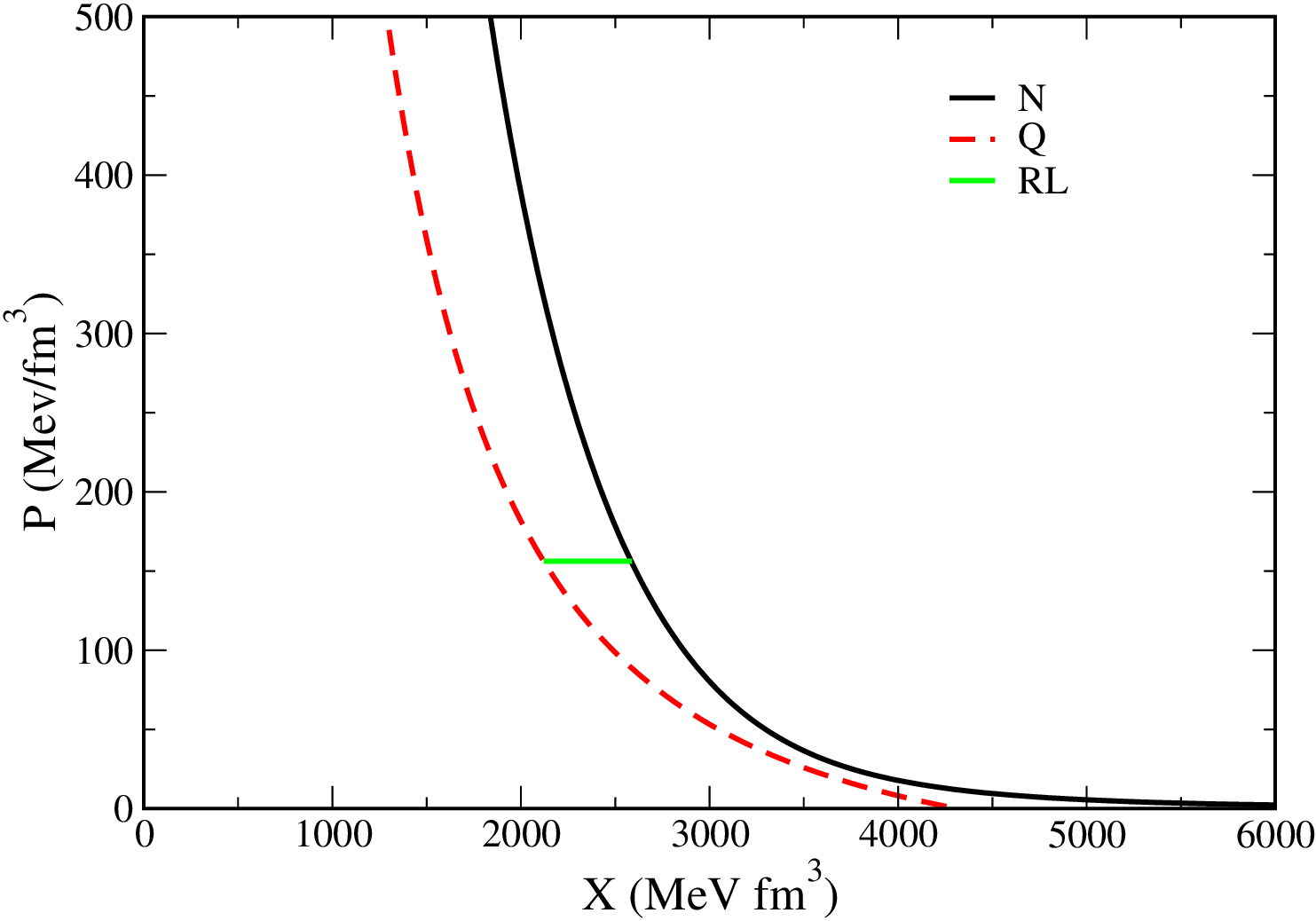}
\caption{(Color online) The Poisson's adiabats plotted in the $X-P$ plane both for NM and QM (Q for QM). 
straight green line (RL) indicate baryon density and energy density jumps for the
equilibrium PT. The graphs are shown for magnetic field induced EoS.}
\label{taub1}
\end{figure}
 
To describe the QM, we use simple MIT bag model \cite{chodos}. The inclusion of the quark interaction in this basic model makes it possible to satisfy the present mass bound. 
The grand potential of the model is given by
\begin{equation}
 \Omega_Q=\sum_i \Omega_i +\frac{\mu^4}{108\pi^2}(1-a_4)+B,
\end{equation}
where $i$ stands for quarks and leptons, $\Omega_i$ signifies the potential for species $i$ and
$B$ is the bag constant. The second term is for the interaction of quarks. $\mu$ is the baryon chemical potential and 
$a_4$ is the quark interaction parameter, varied between 1 (no interaction) and 0 (full interaction). 
We have only two quark species the $u$ and $d$ quarks. The masses of the $u$ and $d$ quarks are $5$ and $10$
MeV respectively. We choose the values of $B^{1/4}=140$ MeV and $a_4=0.56$. 
We have chosen such a parameter setting because we wanted to have PT happening beyond the saturation density. The PT from nuclear matter to quark matter
is usually a two-step process. In the first step, the NM is converted to $2$- flavor quark matter following the hydrodynamic conservation condition.
In the next step, the $2$- flavor matter is converted to a $3$- flavor stable quark matter via weak interaction, which is a slow process. 
As we are dealing with only the first process here, our matter is two flavor matter. All our results are obtained employing $2$- flavor matter properties.
However, for comparison, we have also shown the mass-radius curve for 3-flavour pure QS with s quark having mass of $100$ MeV.

For magnetars, since the magnetic field is very high, it is likely to affect the EoS of the stars. 
The detail of the calculation is similar to that of Mallick \& Sinha \cite{mallick-sinha}, and for brevity, we only give the overall details here. For a magnetic field in 
the $z$-direction, the motion of the charged particles are Landau quantized in 
the $x-y$ plane, and therefore the energy in the n$^{th}$ Landau level is given by 
\begin{equation}
E_n =  \sqrt{p_z^2 + m^2 + 2 n e |Q|B},
\end{equation}
where $p_z$ is the $z$-component of momentum, $B$ is the strength of the magnetic field, $m$ is the mass of the particle, $n$ is the principal quantum number and
$e|Q|$ is the charge of the particle in terms of electronic charge.
The magnetic field similarly modifies the quark matter. The details are given in ref. \cite{mallick-sinha} and we do not repeat them here.
In our calculation we use a simple phenomenological
density dependent magnetic field profile given by 
\cite{chakrabarty,mallick-deform,dexheimer}, and is parametrized as
\begin{equation}
{B}(n_b)={B}_s+B_0\left\{1-e^{-\alpha \left(\frac {n_b}{n_0} \right)^\gamma}\right\},
\label{mag-vary}
\end{equation}
with $B_s$ being the surface magnetic field and $B_0$ is the field at infinitely high density. The surface field is assumed to be $10^{15}$ G and central field is $3 \times 10^{18}$.
We assume $\alpha = 0.01$ and $\gamma = 2$, which is a gentle variation of the magnetic field inside the star. For such a variation the magnetic field at the center of the NS 
is about $1.7 \times 10^{18}$ G or in Lorentz-Heaviside unit it is about $1.2 \times 10^5$ MeV$^2$.

The maximum central field that a star can support is of the order of few times $10^{18} G$. A full general relativistic treatment was first done by Bocquet et al. \cite{bocquet}
and was further developed by Cardall et al. \cite{cardall} where they solved the full Einstein-Maxwell equation and found that if the magnetic field is few times $10^5$ giga Tesla
($10^{18}$ G), the shape of the star becomes toroidal. 
However, such shape of a neutron star is difficult to obtain, and not all numerical code can handle such extreme NS configuration. In most of the recent papers 
\cite{debarati,schramm, veronica} the maximum magnetic field at the center of the star is assumed to be $10^{18}$ G. In our work we have performed calculation with such 
extreme magnetic field strength at the center of the star ($1.7 \times 10^{18}$ G). 

The EoS of the NM and QM are plotted in fig \ref{eos}a. In the same figure, we also plot the magnetic field induced EoS.
We find that both the NM and QM EoS gets softer due to the magnetic field.
The equilibrium PT point between the two phases can be calculated by plotting the pressure of the two phases as a function of chemical potential.  
The point where the two curves cross gives us the PT point. Below the crossing point, the matter is hadronic and above it is quark (as shown in fig \ref{eos}b). 
We find that even for magnetic induced NM and QM EoS the crossing point does not change much.
The PT is implemented assuming Maxwell's construction. The Tolman-Oppenheimer-Volkoff (TOV) equations are solved to obtain the star sequence for the hybrid stars (fig \ref{mr}). 
The low mass stars are all pure hadronic, however, once the central 
density crosses the threshold value, QM starts to appear in their 
core and we obtain a separate branch.
It is easy to observe that the PLZ model generates quite massive stars, the maximum being 2.4 $M_{\odot}$ with a radius of about $11.5$ km. 
The hybrid stars are less massive than the pure NS as they have quark matter inside then. The pure QSs are more compact and gives completely different curves. 
{\bf The curves for the magnetic field induced stars (both neutron and quark) are drawn solving the Einstein field equations in the presence of magnetic field. In section 3, 
we give the essential steps of our calculation in detail.}

The EoS and the equilibrium PT can also be represented in the form of Poisson adiabats as shown in fig \ref{taub1}. The pressure is plotted as a function of 
parameter $X$ given by$X \equiv (\epsilon + p)/ n_b^2$. The straight horizontal line connecting the two phases represents equilibrium PT. 
As we move towards the star core the pressure increases whereas the value of X decreases. The value of X of NM is larger than that of QM because for the same value of 
pressure the density of QM is higher. 
In fig \ref{taub1}, there is only equilibrium PT, and as QM EoS is less stiff than the NM EoS, the Taub adiabat curve lies on the left of nuclear adiabat. 
The Rayleigh line (RL) is also horizontal meaning equilibrium PT, 
where there is a density jump but no pressure jump. If the QM EoS is to be more stiff in the high-density regime (varying the bag constant and/or the coupling $a_4$ term), 
then one can have the QM Taub adiabat on the right of the NM curve as obtained by Furusawa et al. \cite {furusawa1,furusawa2}. 
However, for standard cases, it is highly unlikely.

During the PT there is a jump in the value of X, which becomes stronger at larger densities.
The equilibrium PT is difficult in old cold pulsars unless there is some sudden fluctuation in the thermodynamic quantities which can grow to give a 
step-like feature. We assume that such a step-like discontinuity generated near the star center and which propagates outwards bringing about a PT. The 
PT front burns the NM and leaves behind a compressed quark core. At relatively low-density region the discontinuity diminishes, and PT fronts stop.
It is also assumed that the discontinuity happens only in a very thin layer in comparison to the star radius.

\section{Magnetic field on the star structure}\label{magnetic field}
In the present work, our main aim is to study the effect of strong magnetic fields in the PT of compact stars. {\bf Such a strong magnetic field can have some effect on 
the mass-radius relationship of the star.} The details of the calculation can be found in our previous paper 
\cite{mallick-deform}. Here we only mention the basic details and study their effect on the given stars sequences. 
In this work, we have neglected the effect due to magnetization. The magnetization effect becomes significant only if the central field is greater than $10^{19}$ G, by which 
time the star becomes unstable \cite{sinha}.
The deformation of the star mainly arises due to non-uniform 
magnetic pressure. In the rest frame of the fluid, the magnetic field is in the z-direction, 
the energy density and pressure are given by 
\begin{eqnarray}
 \varepsilon=\varepsilon_m+\frac{B^2}{8\pi}; \\
 P_{\perp}=P_m+\frac{B^2}{8\pi}; \\
 P_{\parallel}=P_m-\frac{B^2}{8\pi},
\end{eqnarray}
where $\varepsilon$ is the total energy density, $\varepsilon_m$ is the matter-energy density and $\frac{B^2}{8\pi}$
is the magnetic stress. $P_{\perp}$ and $P_{\parallel}$ are the perpendicular and parallel components
of the total pressure concerning the magnetic field. $P_m$ is the matter pressure.

The total pressure in both directions can be written as a single equation in terms of spherical harmonics
\begin{eqnarray}
 P=P_m+[p_0+p_2P_2(cos\theta)],
\end{eqnarray}
where $p_0=\frac{B^2}{3.8\pi}$ is the monopole contribution and $p_2=-\frac{4B^2}{3.8\pi}$ the quadrupole contribution of the
magnetic pressure. $P_2(cos\theta)$
is the second-order Legendre polynomial and is defined as
$P_2(cos\theta ) = 1/2 (3cos^2\theta -1)$, where $\theta$ is the polar angle with respect to the direction of magnetic field.

Similarly the metric describing a axially symmetric star can formulated as a multipole expansion
\begin{eqnarray}
ds^2=-e^{\nu(r)}[1+2(h_0(r)+h_2(r)P_2(cos\theta))]dt^2 \\ \nonumber
      +e^{\lambda(r)}[1+\frac{e^{\lambda(r)}}{r}(m_0(r)+m_2(r)P_2(cos\theta))]dr^2 \\ \nonumber
      +r^2[1+2k_2(r)P_2(cos\theta)](d\theta^2+sin^2\theta d\phi^2),
\end{eqnarray}
where $h_0,h_2,m_0,m_2,k_2$ are the corrections up to second order.

The Einstein field equations are used to find the metric potentials in
terms of the perturbed pressure and hence can be solved to calculate the mass modification and axial deformation.
To solve the Einstein equation we use the above mentioned density-dependent magnetic field profile of the star (eqn. \ref{mag-vary}). 
{\bf This simple approach ensures a physical situation that a non-uniform magnetic field is present in the star. The model assumes that the magnetic field at the center
of the star can be several order of magnitude larger than the surface. The anisotropic magnetic pressure generates an excess mass of the star and also produces a significant 
deformation. In the equatorial direction, magnetic pressure adds to the matter pressure causing the equator to bulge, whereas in the polar direction the magnetic pressure 
reduces the matter pressure causing the pole to compress. The star, therefore, takes the shape of an oblate spheroid.}

Using the given prescription, we calculate the stars sequence for NS and HS. In fig \ref{mr} we find that such strong magnetic fields significantly changes the mass-radius nature of 
the curves. 
The effect of magnetic field on the EoS makes it softer which will eventually reduce the maximum mass of the star. However, the magnetic force on the TOV equation tends to increase 
the mass of the star. Ultimately, by the combined action of these two effects the maximum mass of the star does not change much, however, it has a considerable impact on the stars 
radius.

\section{Fluid dynamic conservation conditions}
The differential form of energy-momentum conservation law for a fluid dynamical system is given by 
\begin{equation}
D_\mu T^{\mu\nu}=0,
\end{equation}
where
\begin{equation}
  T^{\mu\nu}=wu^{\mu}u^{\nu}-pg^{\mu\nu}.
\end{equation}
$w$ is the enthalpy ($w=\epsilon+p$), $u^{\mu}=(\gamma,\gamma v)$ is the 
normalized 4-velocity of the fluid and $\gamma$ is the Lorentz factor. $g^{\mu\nu}$ is the 
metric tensor chosen as $(1,-1,-1,-1)$ using standard flat space-time convention. Along with this the 
baryon number is also conserved for an isolated system such as CSs. 
The conservation laws can also be realized in the form of discontinuous hydrodynamical flow usually in shock waves.
We assume that the PT happens as a single discontinuity front propagates separating the two phases. Therefore we denote $``h''$ as the initial state ahead of the shock (NM)
front and $``q''$ as the final state behind the shock (QM).

Across the front, the two phases are related via the energy-momentum and baryon number conservation. 
The relativistic conservation conditions for the space-like (SL) and time-like (TL) shocks are derived from the above-generalized equations \cite{mallick-tl,taub,csernai}.

a. Space-like
\begin{eqnarray}
 w_h\gamma_h^2v_h=w_q\gamma_q^2v_q ;\\
 w_h\gamma_h^2v_h^2+p_h=w_q\gamma_q^2v_q^2+p_q ;\\
 n_hv_h\gamma_h=n_qv_q\gamma_q.
\end{eqnarray}

b. Time-like
\begin{eqnarray}
  w_h\gamma_h^2-p_h=w_q\gamma_q^2-p_q ;\\
  w_h\gamma_h^2v_h=w_q\gamma_q^2v_q ;\\
  n_h\gamma_h=n_q\gamma_q.
\end{eqnarray}

However, when intense magnetic fields are present, the conservation condition gets modified. It now has both matter and magnetic contributions \cite{mallick-tl}.
Infinitely conducting fluid assumption makes the electric field to disappear. Also, the conservation is solved in a particular frame called HT frame 
\cite{hoffmann} where the fluid flows along the magnetic lines, and there are no $\overrightarrow{u}\times \overrightarrow{B}$ electric fields. In this framework, 
the magnetic field and the matter velocities are aligned. We assume that $x$-direction is normal to the shock plane. The magnetic field is constant and lies in the $x-y$ plane. Therefore the velocities and the magnetic fields are given by $v_x$ and $v_y$ and by 
$B_x$ and $B_y$ respectively. The angle between the magnetic field and the shock normal in the HT frame is denoted by $\theta_a$ ($\theta_i$ the incidence angle and 
$\theta_r$ the reflected angle).

Therefore the conservation conditions now read as 

a. Space-like
\begin{eqnarray}
 w_h\gamma_h^2v_{hx}=w_q\gamma_q^2v_{qx};\\
 w_h\gamma_h^2v_{hx}^2+p_h+\frac{B_{hy}^2}{8\pi}=w_q\gamma_q^2v_{qx}^2+p_q+\frac{B_{qy}^2}{8\pi};\\
 w_h\gamma_h^2v_{hx}v_{hy}-\frac{B_{hx}B_{hy}}{4\pi}=w_q\gamma_q^2v_{qx}v_{qy}-\frac{B_{qx}B_{qy}}{4\pi};\\
 n_hv_{hx}\gamma_h=n_qv_{qx}\gamma_q.
\end{eqnarray}

b. Time-like
\begin{eqnarray}
 w_h\gamma_h^2-p_h+\frac{B_{hy}^2}{8\pi}=w_q\gamma_q^2-p_q+\frac{B_{qy}^2}{8\pi};\\
 w_h\gamma_h^2v_{hx}=w_q\gamma_q^2v_{qx};\\
 w_h\gamma_h^2v_{hy}=w_q\gamma_q^2v_{qy};\\
 n_h\gamma_h=n_q\gamma_q.
\end{eqnarray}

For the HT frame we also have 
\begin{eqnarray}
 \frac{v_{hy}}{v_{hx}}=\frac{B_{hy}}{B_{hx}}\equiv \tan\theta_i , \\
 \frac{v_{qy}}{v_{qx}}=\frac{B_{qy}}{B_{qx}}\equiv \tan\theta_{r}.
 \label{mag-theta}
\end{eqnarray}
The assumption of infinite conductivity gives the electric field to be zero.
The Maxwell equation of no monopoles $\nabla \cdot \overrightarrow{B}=0$ gives
\begin{equation}
 B_{hx}=B_{qx}.
\end{equation}
Substituting this in eqn. 26 and 27, we have $B_{qy}/B_{hy}=\tan\theta_r /\tan\theta_i$. Usually, $\theta_r \neq \theta_i$ and therefore the magnetic field across
the PT is discontinious.

The TL conservation conditions lead to some exciting results in the HT frame, which can be obtained analytically.
Dividing eqn. 24 by eqn. 23, we get
\begin{equation}
 \frac{v_{hy}}{v_{hx}}=\frac{v_{qy}}{v_{qx}}.
\end{equation}
Combining this with eqn. 26 and eqn. 27, we have
\begin{equation}
 \frac{B_{hy}}{B_{hx}}=\frac{B_{qy}}{B_{qx}}.
\end{equation}
But eqn. 28 says $B_{hx}=B_{qx}$, therefore we have $B_{hy}=B_{qy}$.

Therefore eqn. 22 now becomes 
\begin{equation}
 w_h\gamma_h^2-p_h=w_q\gamma_q^2-p_q,
\end{equation}
which is same as the non-magnetic case.

Eqn. 23 and 24 can be combined in a single equation 
\begin{equation}
 w_h\gamma_h^2v_{h}=w_q\gamma_q^2v_{q}.
\end{equation}

Therefore, the TL conservation equation remains the same as the nonmagnetic case. The magnetic field does not affect the TL PT or discontinuity.
In our previous work \cite{mallick-tl} we found somewhat similar result numerically, but here we can work them out even analytically. 
The conservation condition is such that the magnetic field does not affect TL shocks and only matter properties govern them.
In this work, we will, therefore, discuss only SL shocks.

\section{Results}
Fluctuation of the thermodynamic quantities at the center of the star starts the PT. The PT will depend on the process being exothermic or endothermic.
At the center of the star first there is a deconfinement transition, and then there is conversion to stable QM ($3$- flavor). At a certain point as the matter converts from 
NM to QM ($2$- flavor) there is a sharp change in the thermodynamic variables (like density, pressure, etc.). From here on QM will imply $2-$ flavor QM unless stated otherwise.
The propagation of the PT front depends on the energy difference between the NM and QM. If the energy of the NM is greater than that of the QM (at fixed number density), 
then the conversion is exothermic, 
and shock-like features can develop. However, the energy difference depends on the EoS of 
NM and QM, the baryon density at which the PT is taking place and also on the velocity of the shock front. The above conservation conditions are written in the rest frame of the 
conversion front. We solve our problem in this frame and switch to the global frame where QM is at rest.  In our calculation, for the front rest frame,
NM velocity is represented as $v_h$ and QM velocity as $v_q$. In the global frame, NM velocity is given by $v_n$ and front velocity as $v_f$.
In this global frame the NM moves toward the
center with velocity $v_n = (v_h - v_q )/(1 - v_h v_q )$. The front velocity near the center can assumed 
to be $v_f=-v_q$, where $v_h$ and $v_q$ are the quantities in the HT frame.

The equilibrium PT from NM to QM happens at around $3.2$ times nuclear saturation density for our chosen set of EoSs. Therefore we want to examine the shock-induced PT around this 
point for comparison. Therefore, we choose shock induced PT happening at $3$ times and $4$ times saturation density $\rho_s$. The magnetic field at these points can be calculated 
from our chosen magnetic field profile. The magnetic field $B$ at $3 \rho_s$ is $6 \times 10^{17}$ G ($4.3 \times 10^4$ MeV$^2$) and at $4 \rho_s$ is $1 \times 10^{18}$ G 
($7 \times 10^4$ MeV$^2$).

\begin{figure}
\centering
\includegraphics[width = 3.4in]{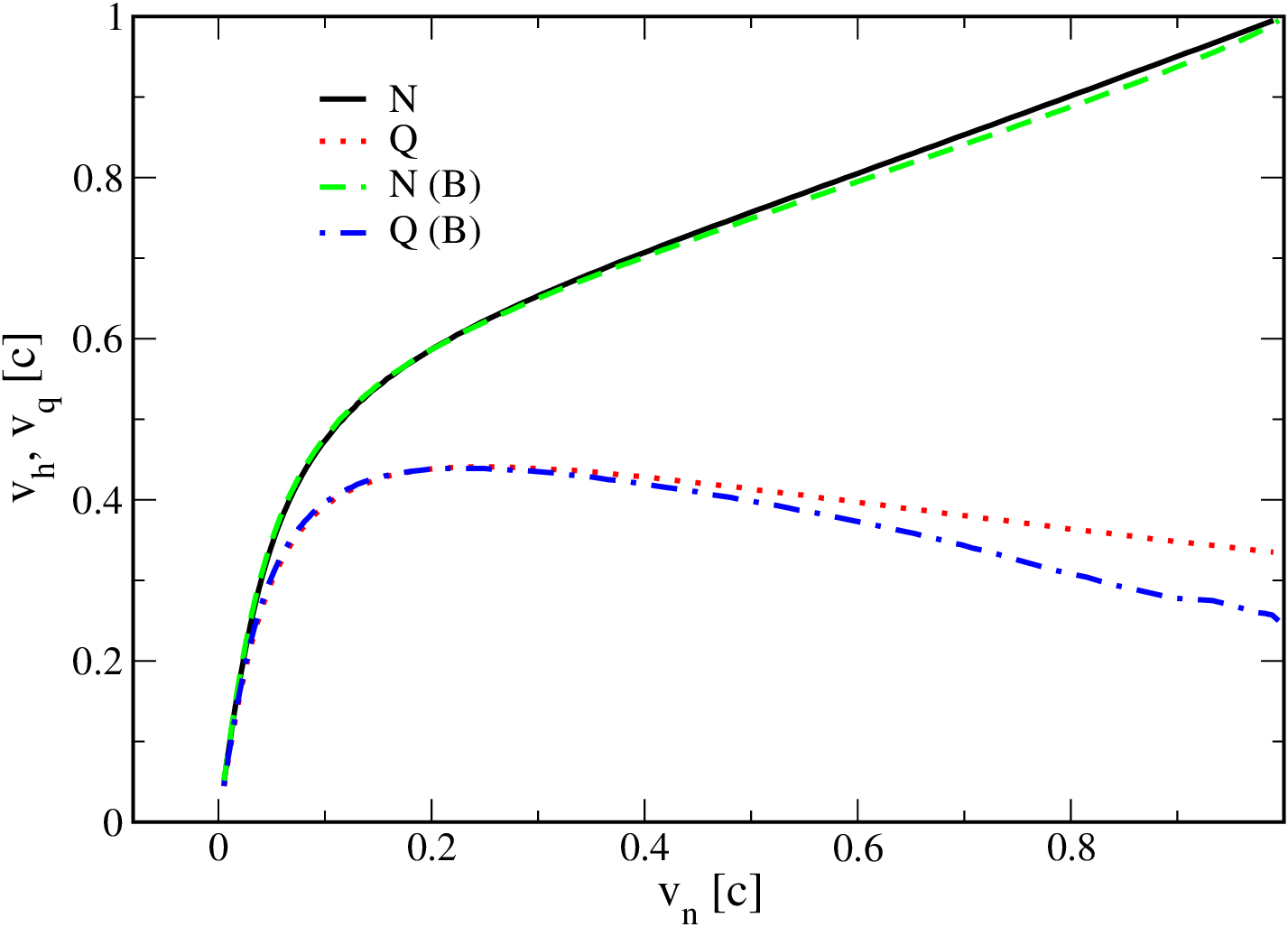}
\caption{(Color online) The variation of $v_h$ and $v_q$ as a function of $v_n$ is shown. a) Solid (without B) and dashed (with B) curve corresponds 
to $v_h$ and dotted (without B) and dash-dotted (with B) curve corresponds to $v_q$. Curves are shown for $\rho_b=4\rho_s$.}
\label{vel1}
\end{figure}

In this calculation, we will see the evolution of our relevant parameters regarding $v_n$ and nuclear density $\rho_b$. 
For such analysis first we have to know how $v_h$ and $v_f$ varies with $v_n$. The velocity variation is shown in fig \ref{vel1}. We see that initially as $v_n$ increases 
from $0$ to $0.1$ there is a 
sharp rise in $v_h$ from $0$ to $0.5$ after which the slope of $v_h$ decreases and gradually 
goes to $1$ as $v_n$ reaches $1$. However, the value of $v_h$ is always greater than $v_n$. The difference is high at lower values of velocity and decreases as velocity increases. 
On the other hand, if we see the variation of $v_f$ (which is the same as $v_q$
only the direction changes), we find that initially $v_f$ rises rapidly to attain a maximum value of $0.44$ at $v_n=0.2$ and from there it decreases slowly to attain value 
$1/3$ as both $v_n$ and $v_f$ goes to $1$. Near the center of the star, the velocity of the incoming matter is largest. 
As the shock wave propagates outwards from the center, the incoming matter velocity decreases but the front velocity increases gradually. 
However, after $v_n$ becomes 
less than $0.2$, the front velocity drops rather quickly and vanishes at $v_n \rightarrow 0$. It is interesting to note that the conservation conditions act in such a 
way that even without any dissipation mechanism there is some deceleration which drives the front velocity to zero at some point inside the star, corresponding to an equilibrium configuration with the static phase boundary. The $v_h$ and $v_q$ with magnetic field follows the non-magnetic curve closely, but their value at any $v_n$ is slightly smaller than their nonmagnetic counterpart. The magnetic field slows front velocity or the speed of the conversion. Thus the PT happens slowly for magnetars than in normal NSs.

\begin{figure}
\subfloat[]{\includegraphics[width = 2.4in]{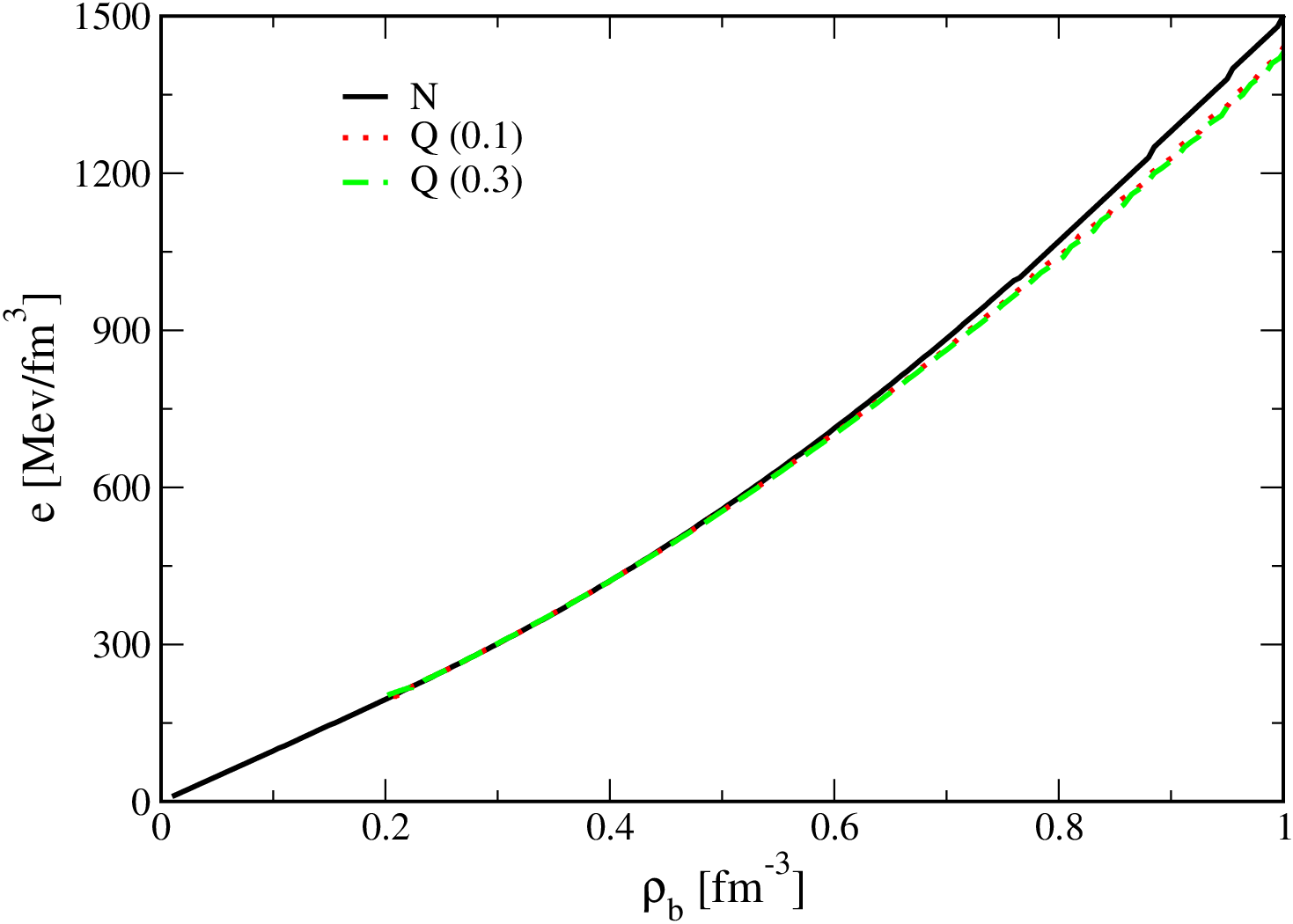}} \quad 
\subfloat[]{\includegraphics[width = 2.4in]{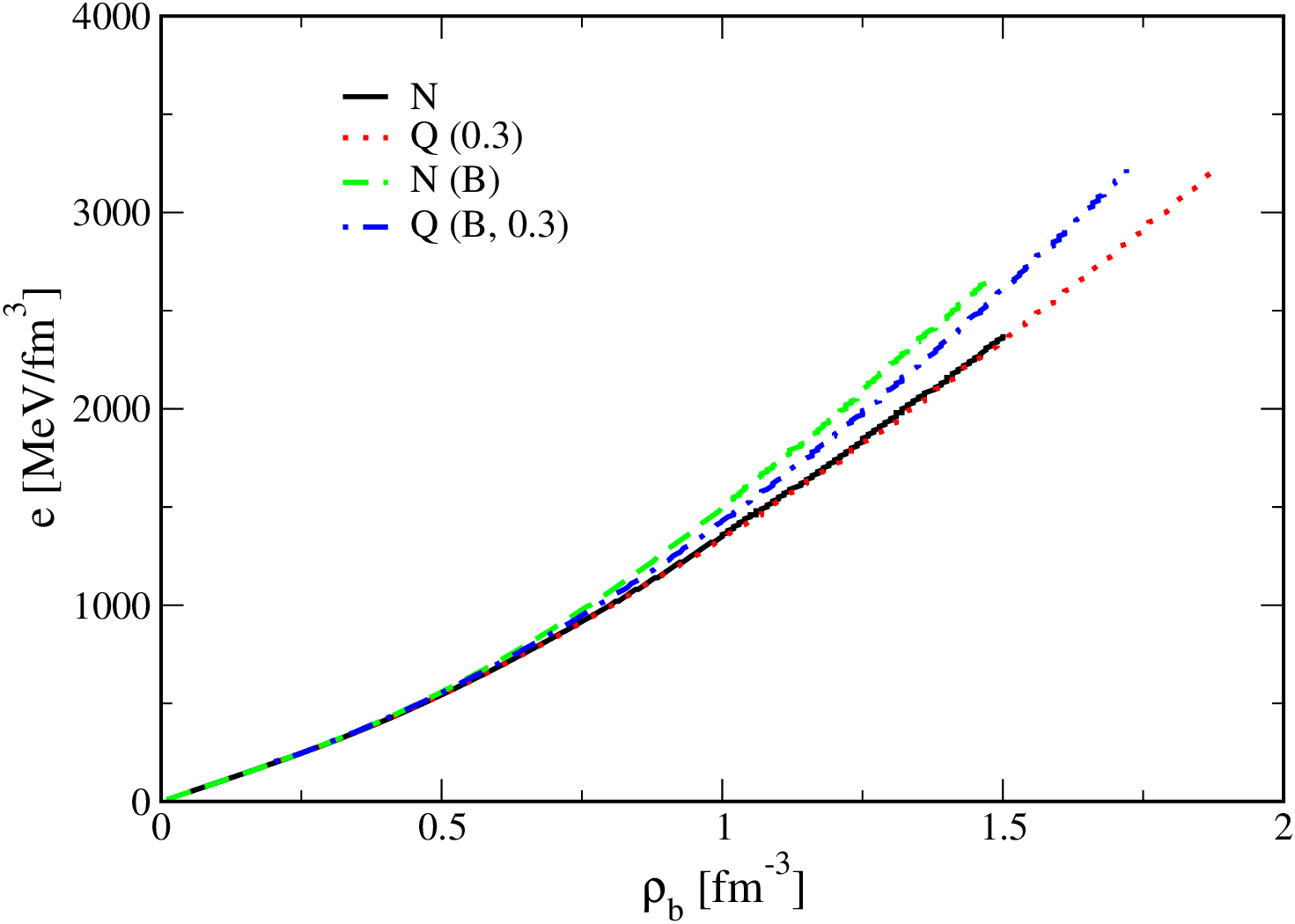}}
\caption{(Color online) The energy density of the NP and QP as a function of baryon density are shown. a) Curves without magnetic field contribution, 
plotted for two given values of $v_h$, $0.1$ and $0.3$.
b) Curves with contribution from magnetic field are shown with $v_h=0.3$. }
\label{exo}
\end{figure}

If there is a sudden fluctuation of matter at high density, then the PT is no longer an equilibrium PT. Such delayed PT can occur at higher densities and can be a violent one. 
Such PT would depend primarily on the energy difference between hadronic state and the shocked quark state. In fig \ref{exo}a we see that the energy of the NM is considerably higher than the energy of the QM
(corresponding to a particular number density) only beyond $3 \rho_s$. Therefore the PT beyond this point is an exothermic one, and shock-like features can develop. We also find that at densities below $2\rho_s$ the energy difference between the un-shocked and shocked phases becomes almost equal. Beyond these densities inside the star, it is difficult for the matter to undergo PT. At such densities, it is expected that the dynamical shock-front would decelerate and ultimately stop. However, such analysis can only be done once we do the full dynamic calculation, which is beyond the scope of this article. We have shown curves for two different incoming hadronic velocities ($v_h$), and the behavior of the curves does not change much. In fig \ref{exo}b 
we show similar curves but with contribution from magnetic energy. 
The magnetic energy adds to the matter energies and makes the curves stiffer. However, the PT is still exothermic.

\subsection{magnetic field and tilt of magnetic axis}

\begin{figure}
\subfloat[]{\includegraphics[width = 2.4in]{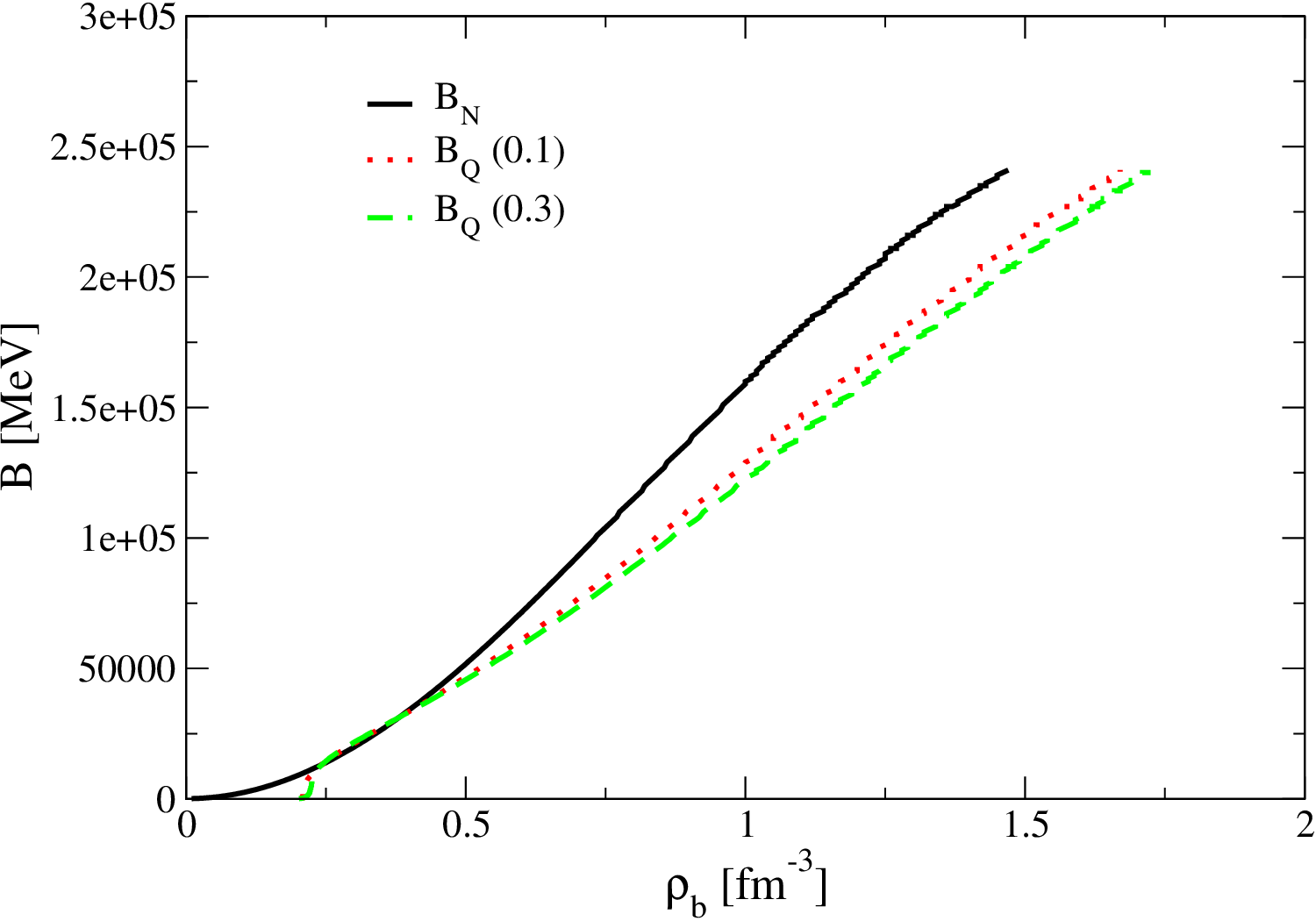}} \quad 
\subfloat[]{\includegraphics[width = 2.4in]{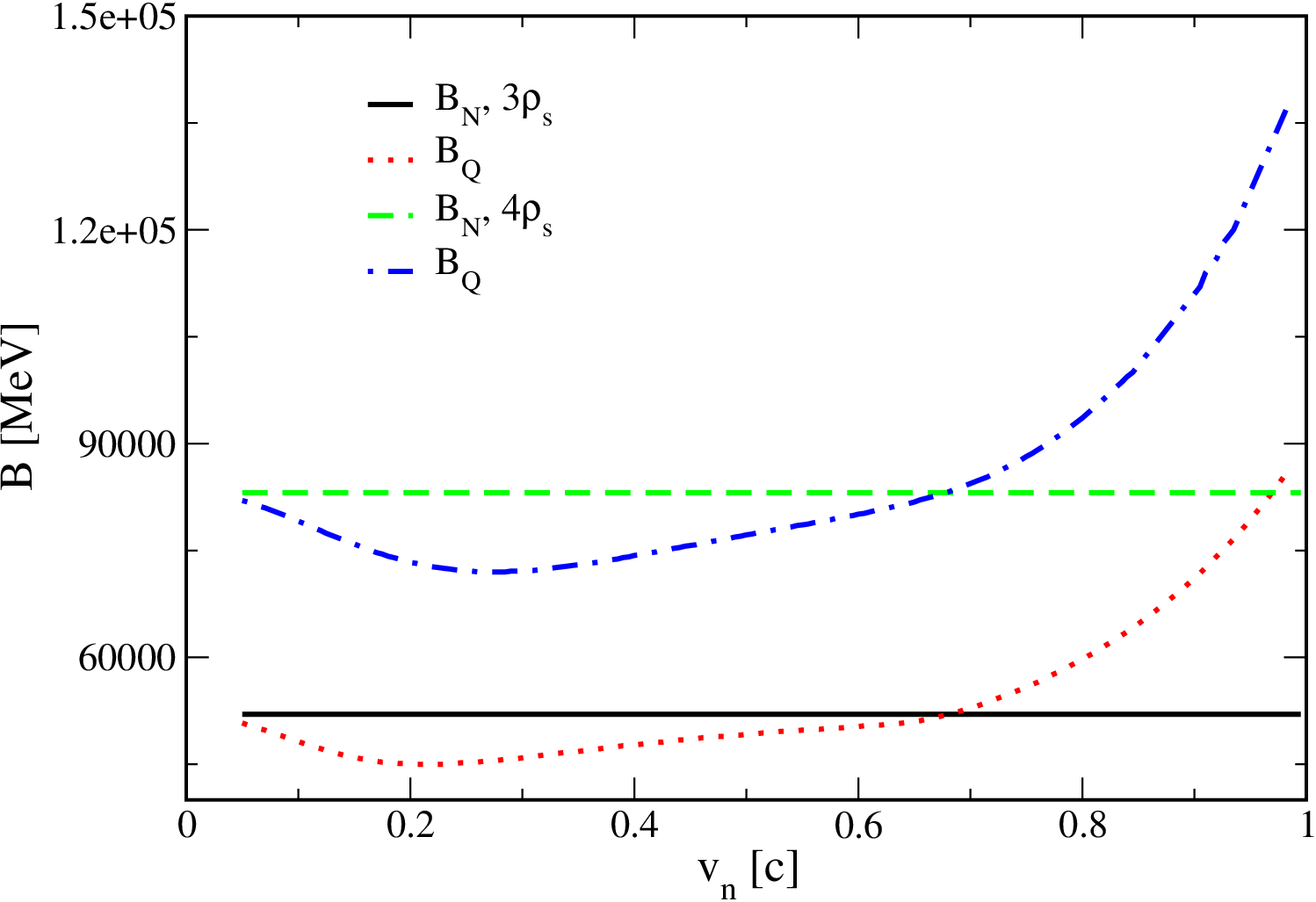}}
\caption{(Color online) Magnetic field in the NP and QP are shown as a function of $\rho_b$ and $v_n$. a) Solid line is for NP and the 
dotted and dashed lines are for QP with $v_h=0.1$ and $v_h=0.3$ respectively. b) Similar set of curves are plotted as a function of $v_n$ for two 
different $\rho_b$, $3\rho_s$ and $4\rho_s$. 
Quantities with subscript $N$ corresponds to NP and with subscript $Q$ corresponds to QP.}
\label{mag-rhov}
\end{figure}

The magnetic field for the the NM is our input. Whereas, the magnetic field for the QM is obtained by solving the conservation conditions.
The strength of the magnetic field in QM will determine whether the resultant star is more or less magnetic. In our previous calculations \cite{mallick-sinha}, we have seen 
that the QS is less magnetic than a NS where the EoS was solely responsible for the magnetic field calculation, however, in our current study, we refine our calculation by 
solving the magneto-hydrostatic conditions (conservation conditions have magnetic field exclusively) with magnetic field induced EoS. 

We have studied the magneto-hydrostatic conservation equations when the magnetic field is at an angle $\theta_i=30^\circ$. In usual pulsars, the magnetic axis (MA) is slightly tilted 
from the body axis.
It has been argued by Flowers \& Ruderman \cite{flowers} that at the birth, the tilt angle is small and grows with time as the stars slow down. Another group \cite{radhakrishnan} 
presented a significantly different picture. However, a recent study proposed \cite{tauris, young,rookyard} that the spin angle is either small (less than $40^{\circ}$) or very large 
(greater than $80^{\circ}$). If the magnetic field is perpendicular to the shock front, then there is no effect as then we will only have $x$-component of the magnetic field 
(no $B_y$ term).  
But from Maxwell's equation (eqn. 28), they are equal in the burnt and unburnt matter. Therefore, for a large tilt angle, the magnetic field will not have much effect on the PT front. 
Thus, the compelling case to study will be the scenario when the magnetic field is less than $40^{\circ}$. For a more conservative approach, we have assumed the magnetic tilt to be $30^\circ$.

\begin{figure}
\centering
\includegraphics[width = 3.5in]{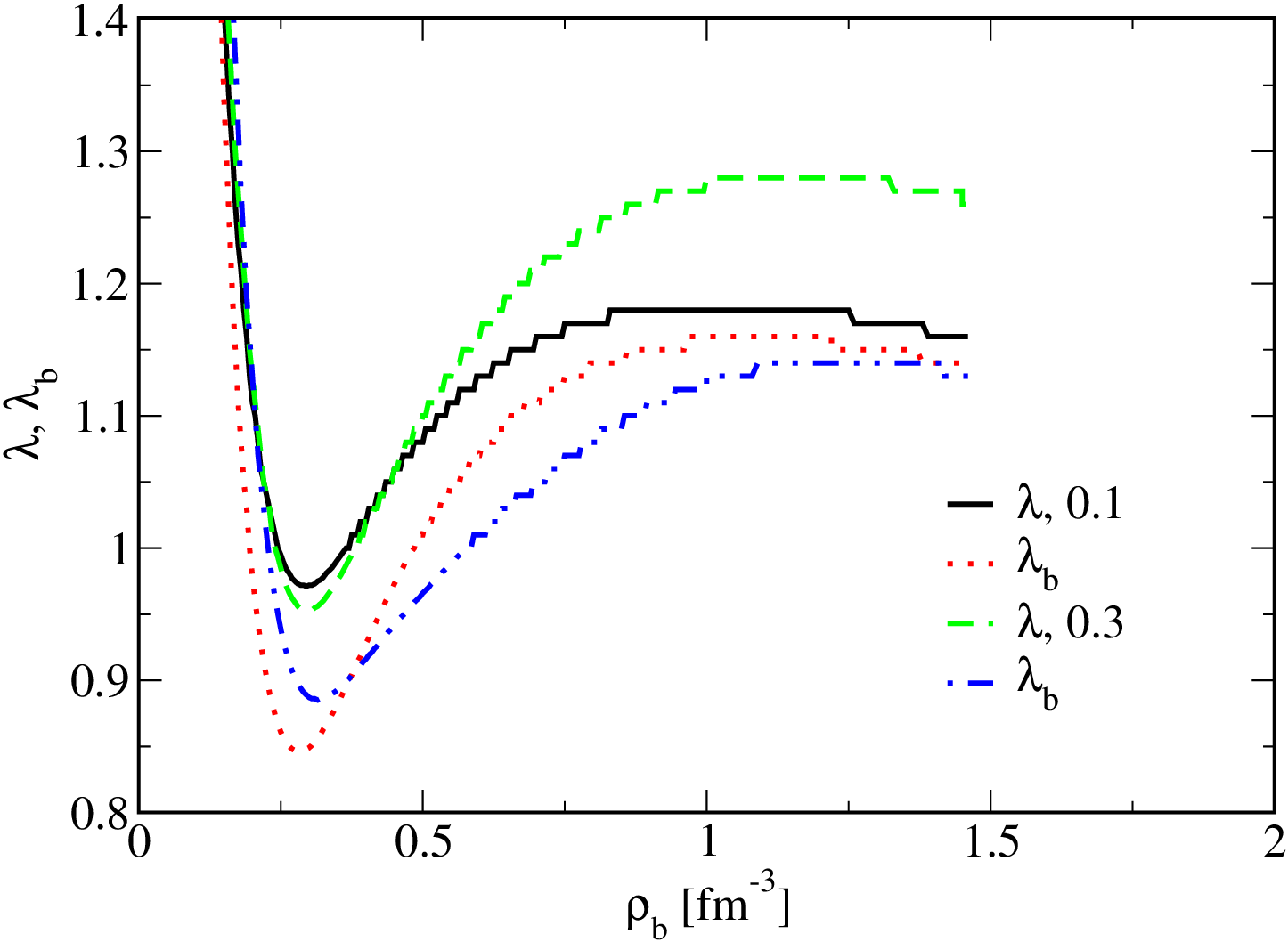}
\caption{(Color online) $\lambda=n_q/n_h$ and $\lambda_b=\lambda. B_q/B_h$ are shown as a function of $\rho_b$ for two values of $v_h$. Black-solid and green-dashed lines represent $\lambda$ curves 
and red-dot and blue-dash-dot curves represent $\lambda_b$.}
\label{field}
\end{figure}

In fig \ref{mag-rhov}a and fig \ref{mag-rhov}b, we plot the initial and final magnetic field as a function of density and $v_n$ respectively.  In fig \ref{mag-rhov}a we plot 
the magnetic field as a function of density for two incoming matter velocities, $0.1$ and $0.3$. As evident from the energy diagrams we conclude that the PT would only be 
possible at densities beyond $3$ times the saturation density.
At such high densities, the incoming matter velocities would not be very high, and therefore we choose $v_n$ to be small.
From fig \ref{mag-rhov}a, we find that the magnetic field in the burnt QM is smaller than the magnetic field in the NM at a fixed density. 
As the density increases the magnetic field in the burnt matter becomes much lower. The change in magnetic field across the two phases is about $12 -15\%$, at the core of the stars. 
As the velocity of the incoming matter increases the magnetic field in the QM decreases much further.

Next, we plot the magnetic field on the two sides as a function of $v_n$ in fig \ref{mag-rhov}b, for two densities. We see that initially at smaller values of $v_n$ 
the magnetic field in the QM decreases 
further below that of NM and attains a minimum value at 
around $v_n=0.2-0.3$ (corresponding to $v_h=0.48-0.65$). From there onwards the magnetic field in QM increases and at values greater than $v_n=0.7$ ($v_h=0.85$) 
the magnetic field in the QM 
becomes greater than that of NM (there is a crossing in the curves).
The nature of the curves remains almost the same for both sets of densities. Such behavior may be because at such 
$v_n$ the value of $v_h$ becomes quite high, and then the conservation condition is driven mostly by the matter enthalpy and pressure than that of the magnetic force.

It is difficult from fig \ref{mag-rhov}a to figure out whether at a particular point in QS the magnetic field would be greater than or less than that of an NS. 
It is because fig \ref{mag-rhov}a
only implies that at a particular density the magnetic field in NM is greater than QM. However, if the QS has resulted from a PT then for a particular point 
(radial distance from the center)
the density would also increase. As the magnetic field strength is a function of density, a rise in density implies a rise in the magnetic field. Therefore, 
the ultimate magnetic field strength is
obtained when we take both the effect into account. This has been shown in fig \ref{field}. Here we plot $\lambda$ and $\lambda_b$ as a function of density, with the 
definition
$\lambda=n_q/n_h$ (ratio of densities) and $\lambda_b= \lambda . B_q/B_h$. The curves show that at very low baryon density $\lambda$ is $>>1$. 
It is less than $1$ between $\rho_b= 0.24 - 0.39$ (for $v_n=0.3$) 
and at higher densities, it again becomes greater than $1$. The $\lambda_b$ curve is similar in nature but attains a value less than $1$ only in the range $\rho_b= 0.23 - 0.58$. The value of $\lambda_b < 1$ indicates
that the QS magnetic field strength (at some particular radial point in the star) is less than NS magnetic field strength. Although the magnetic field in the QM is always less than NM for a particular density, still the magnetic field strength of the QS can be higher than NS at very small and very large densities 
(excluding the range $\rho_b= 0.23 - 0.58$). We find the similar behavior of magnetic field strength for $v_n=0.1$.

\begin{figure}
\subfloat[]{\includegraphics[width = 2.4in]{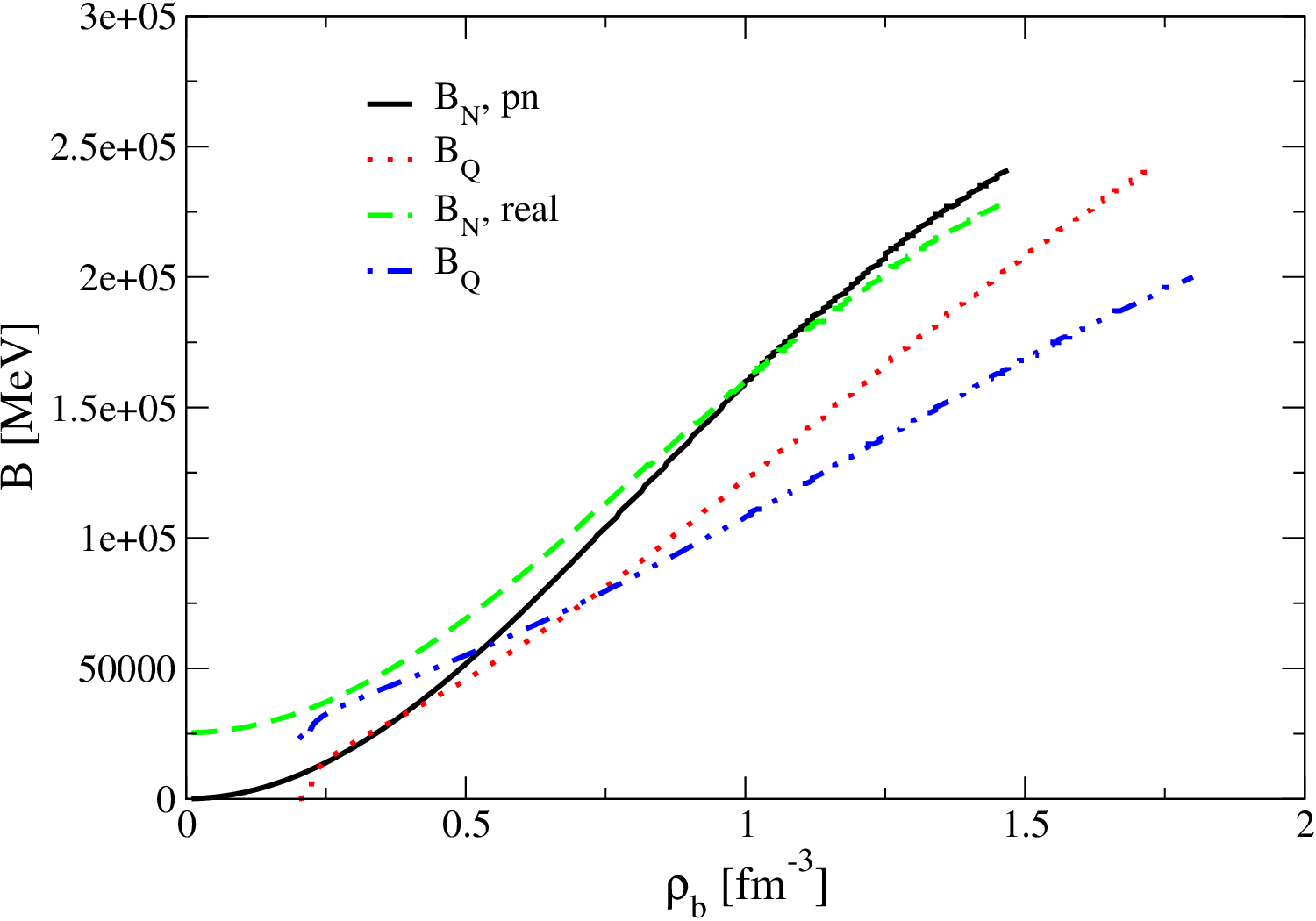}} \quad 
\subfloat[]{\includegraphics[width = 2.4in]{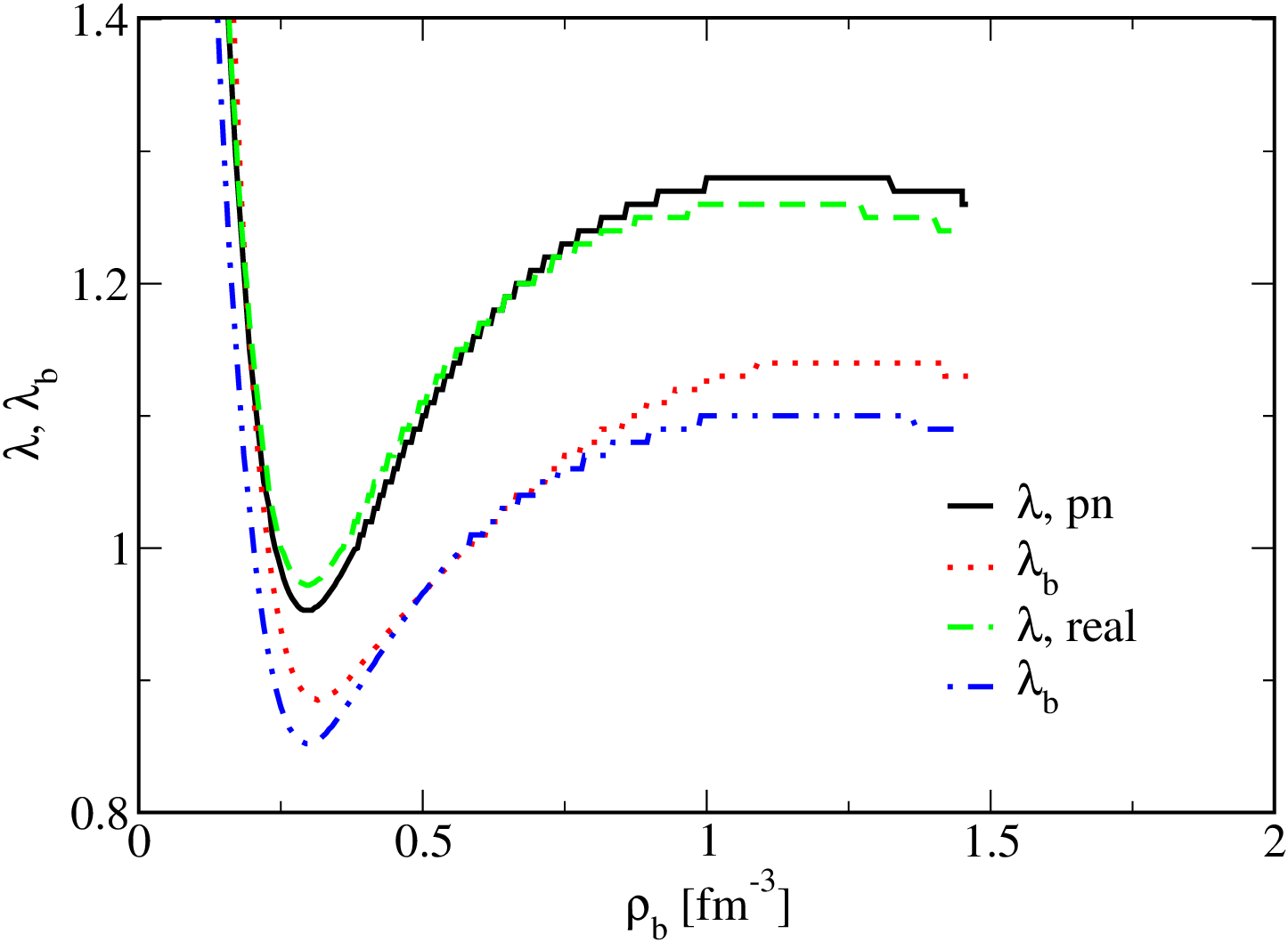}}
\caption{(Color online) a) Magnetic field of NM and QM are shown as a function of baryon density. The script ``pn'' indicates the phenomenological field variation with centre and 
surface field differing
by 3 order of magnitude whereas ``real'' indicates where the surface to centre field strength differing by only 1 order of magnitude.
b) Curves indicates how $\lambda$ and $\lambda_b$ vary as a function of $\rho_b$. The nomenclature remains the same. All the curves a) and b) are shown for $v_h=0.3$. }
\label{real}
\end{figure}

The magnetic field profile that we choose is not consistent with the Einstein-Maxwell equations. Results for a more realistic field variation of one order of magnitude
from the center of the surface is plotted in fig \ref{real} (curves marked ``real''). In this calculation, the surface field is assumed to be $1.6 \times 10^{17}$ G, and the 
center field is $1.6 \times 10^{18}$ G (to compare with previous results) with other parameters remaining the same. The magnetic field strength for QM is much lower than that 
of NM for a fixed density. With this realistic field variation, the magnetic field strength in the QM is reduced by about $25 \%$. However, when we plot the 
$\lambda$ and $\lambda_b$ as a function of $\rho_b$ we find that there is not much change in the region of the star where the magnetic field of QS is less than the magnetic 
field of NS (the density range being $\rho_b= 0.2 - 0.58$).
We, therefore, infer that a realistic magnetic field profile of the magnetic field can have some quantitative changes in the results but the qualitative nature of the 
study remain the same.

\begin{figure}
\subfloat[]{\includegraphics[width = 2.4in]{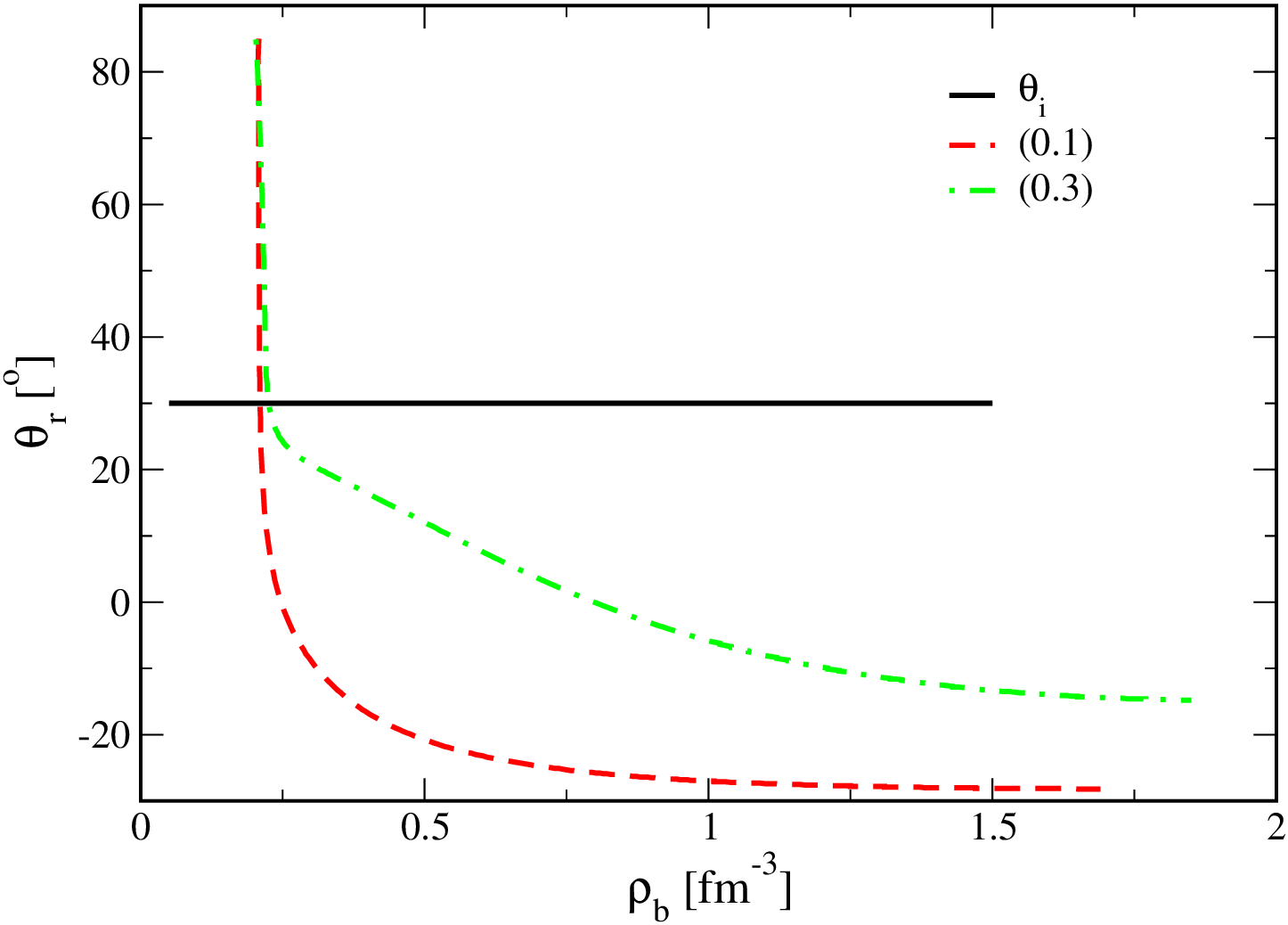}} \quad 
\subfloat[]{\includegraphics[width = 2.4in]{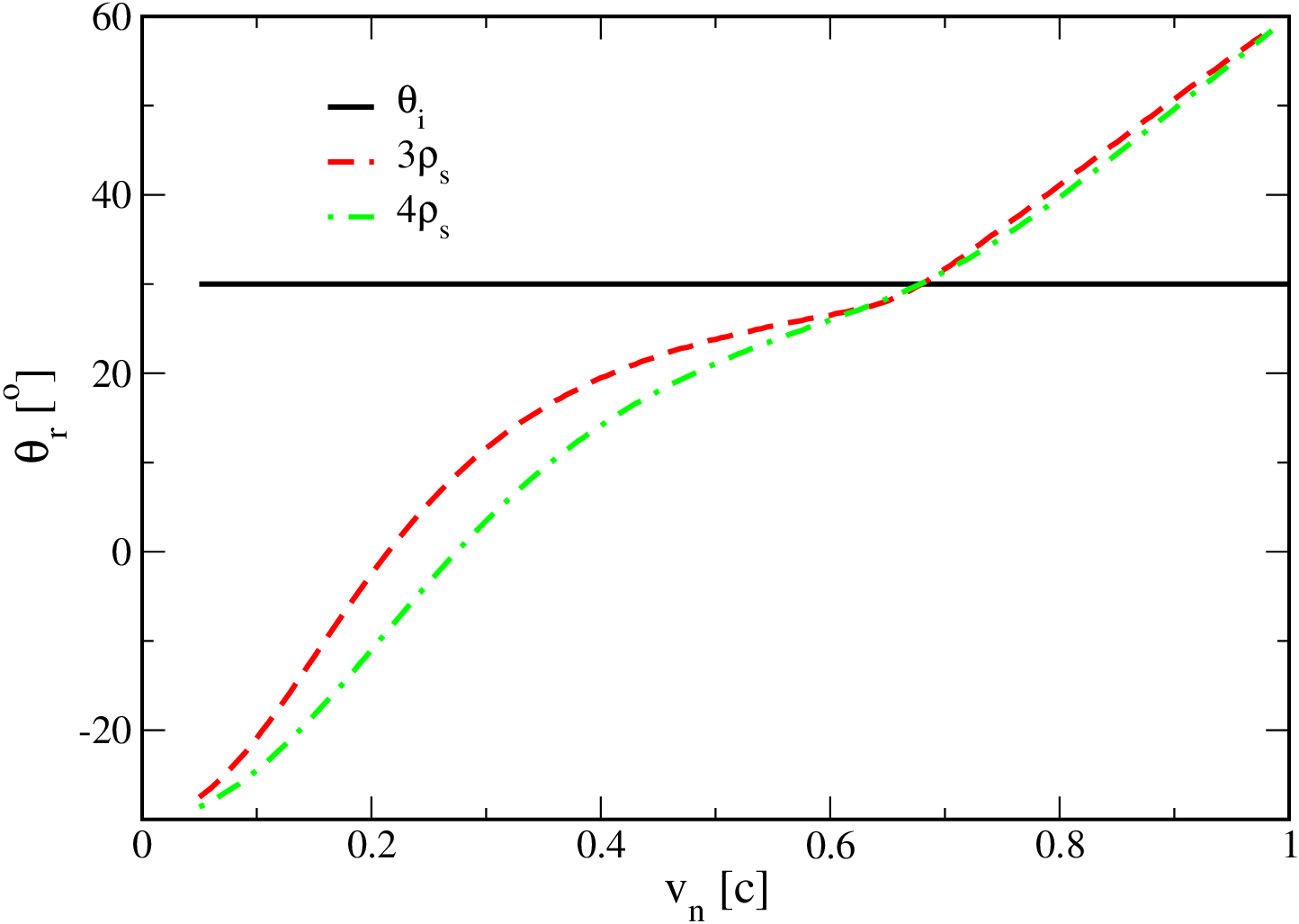}}
\caption{(Color online) Reflected angle ($\theta_r$) is plotted as a function of $\rho_b$ and $v_n$. a) Solid line is for incident angle ($\theta_i$) and the 
dashed and dash-dotted lines are for reflected angle ($\theta_r$) with $v_h=0.1$ and $v_h=0.3$ respectively. b) Similar set of curves are plotted as a function 
of $v_n$ for two different $\rho_b$, $3\rho_s$ and $4\rho_s$.}
\label{theta}
\end{figure}

Another interesting outcome of the phase transition which is only present for MHS calculation is the angle between the magnetic field and shock front. If the star is more or less spherical, then the angle between the body axis and MA will also be the angle between the magnetic field and shock front. Therefore, the incident angle 
$\theta_i$ in the NM is $30^{\circ}$. The angle in the QM is obtained by solving the conservation conditions and is depicted as reflected angle $\theta_r$.
The reflected angle $\theta_r$ changes both with $\rho_b$ and $v_n$. The incident angle $\theta_i$ is always fixed at $30^\circ$ for a star with a magnetic field. In fig \ref{theta}a,
we plot the reflected angle as a function of $\rho_b$ for two incoming matter velocity $v_h$. 
At smaller densities, the reflected angle is large but falls off very fast, and at densities which are of our interest (beyond three times $\rho_s$), the reflected angle is always 
smaller
than incident angle. For small $v_h$ the reflected angle becomes negative. The negative sign in the reflected angle means that the reflected matter velocities and magnetic field are above the plane perpendicular to the shock front (the x-axis). At very higher densities the reflected angle is always negative, and for $v_h=0.1$ it is negative for almost all densities of our interest. However, for $v_h=0.3$ and at some intermediate densities the reflected angle is positive.
So there is a complete change in angular directions at higher densities. Thus, instead of pointing upwards, the matter velocities and magnetic fields point downward in the burnt matter. Such features become clearer in fig \ref{theta}b where we plot $\theta_r$ as a function of $v_n$. We have plotted the curves for two different densities 
($3\rho_s$ and $4\rho_s$). For $3 \rho_s$ we find that as the velocity increases
$\theta_r$ decreases and becomes zero at $v_n=0.21$ ($v_h=0.5$). Beyond that, it increases in the positive direction and goes to about $60^\circ$ at higher velocities. 
It signifies that the outflow velocity and magnetic field in the burnt QM changes direction at higher speeds. The curve for $\rho_b=4\rho_s$ shows almost similar pattern 
only differing numerically. This is a fascinating result as it shows that for violent shocks where the initial incoming matter velocity is high
the resultant QS can have MA tilted in altogether another direction. Previous studies which discuss the evolution of the magnetic tilt axis has calculated its 
development based on the spin frequency or the cooling rates which happen gradually. However, the change in the magnetic tilt brought about by PT is a sudden change and 
can have enormous observational significance. 

\subsection{combustion process}

The variation of matter velocities and the comparison of the burnt and unburnt matter velocities is a valuable tool to understand whether a shock propagation is a detonation or 
a deflagration. If the speed of the burned matter is higher than unburnt matter, the PT is a detonation one, whereas if the velocity of the unburnt matter is higher than burned matter it is a deflagration.
Detonation is very fast burning whereas deflagration is slow combustion. Another way of determining detonation and deflagration is comparing their energy and pressure. 
It can be classified as \\
a) $v_q > v_h, ~ e_q-p_q < e_h - p_h$ detonation. \\
b) $v_q < v_h, ~ e_q-p_q > e_h - p_h$ deflagration.

\begin{figure}
\subfloat[]{\includegraphics[width = 2.4in]{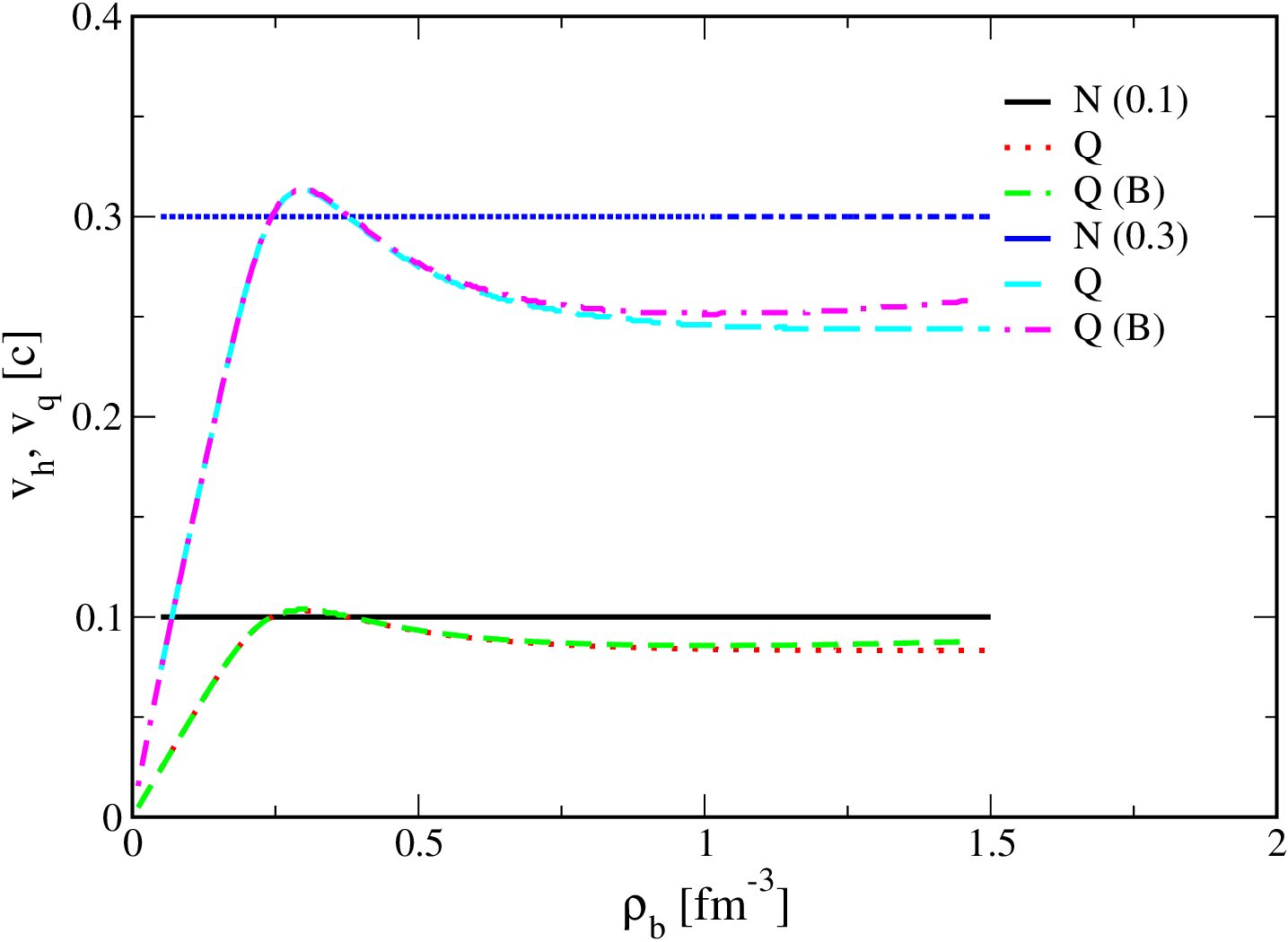}} \quad 
\subfloat[]{\includegraphics[width = 2.4in]{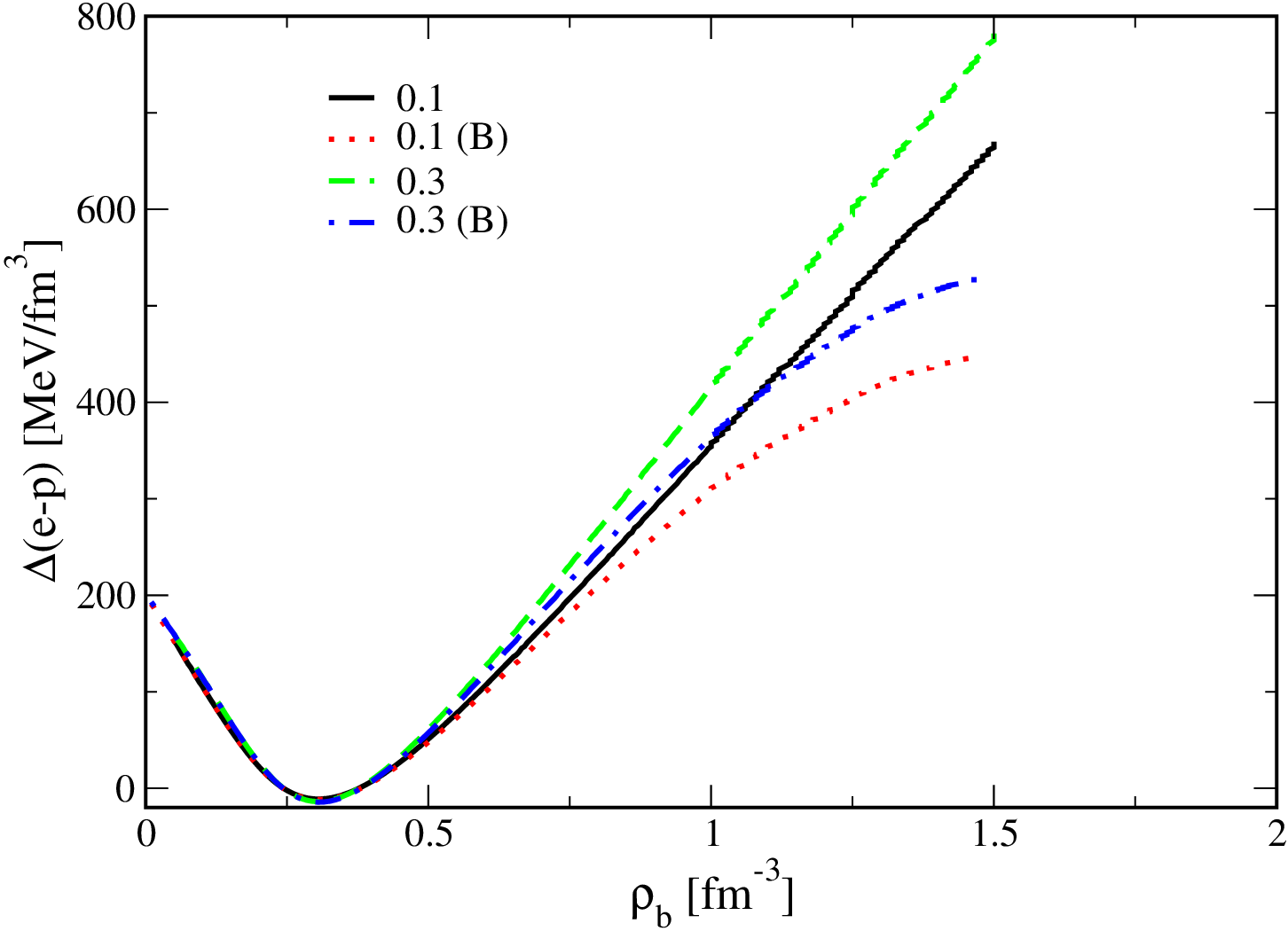}}
\caption{(Color online) a) Curves in this figure drawn to compare $v_h$ and $v_q$ as a function of $\rho_b$. The input $v_h$ for the magnetic and non-magnetic plots are same. 
Solid curve shows $v_h=0.1$ and the two curves below it 
shows the corresponding $v_q$ with (dashed) and without (dotted) magnetic effect. The other family of three is for $v_h=0.3$ and its corresponding $v_q$'s.
b) The difference of (e-p) for the QP and NP ($\Delta (e-p)$) is shown as a function of $\rho_b$. The solid line (without magnetic effect) and the dotted line 
(with magnetic effect) are plotted
for $v_h=0.1$ and the dashed and dash-dotted curves are for $v_h=0.3$.}
\label{deflag}
\end{figure}

In fig \ref{deflag}a we plot $v_h$ and $v_q$ as a function of $\rho_b$.  We have compared it for two values of $v_h$, $0.1$ and $0.3$. We find that for the non-magnetic case at 
very small densities (below two times $\rho_s$) the velocity of the quark matter is greater than the velocity of the nuclear matter. At such densities, the burning process can be a 
detonation one. When we draw similar plot taking into account the magnetic field the quark matter velocity is found to be always smaller than hadronic matter velocity. Therefore, for the magnetic case, the propagation is still a deflagration one.
Such pattern is seen for both the incoming matter velocities. 
Following the condition given for determining detonation and deflagration we plot $(e_q-p_q)-(e_h-p_h)$ (denoted as $\Delta (e-p)$ in the fig \ref{deflag}b) as a function of $\rho_b$. 
The value is always positive, apart from a small window at low densities where it becomes negative. The density range where a detonation process can develop is the same for both 
the plots. The PT is mostly a deflagration type apart from a small window at lower densities. After the small window where it becomes negative the value is always positive 
and increases in density, meaning that the deflagration speeds up at higher densities. For $v_h=0.1$,
at such densities, the magnetic and non-magnetic curve almost overlap, but at higher densities, the magnetic curve lies below the non-magnetic curve. 
The nature of the curves remains the same for all the cases.

\begin{figure}
\subfloat[]{\includegraphics[width = 2.4in]{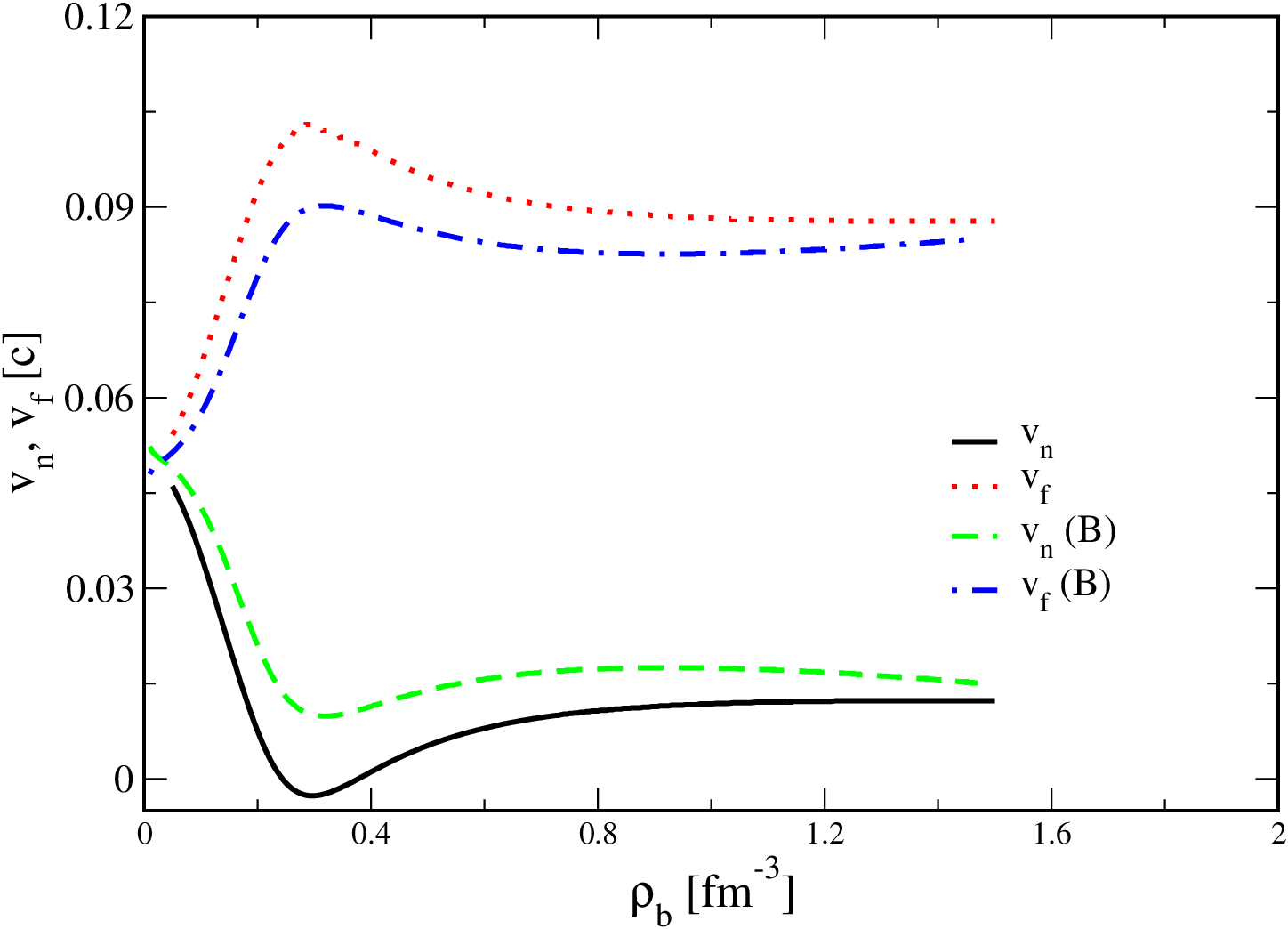}} \quad 
\subfloat[]{\includegraphics[width = 2.4in]{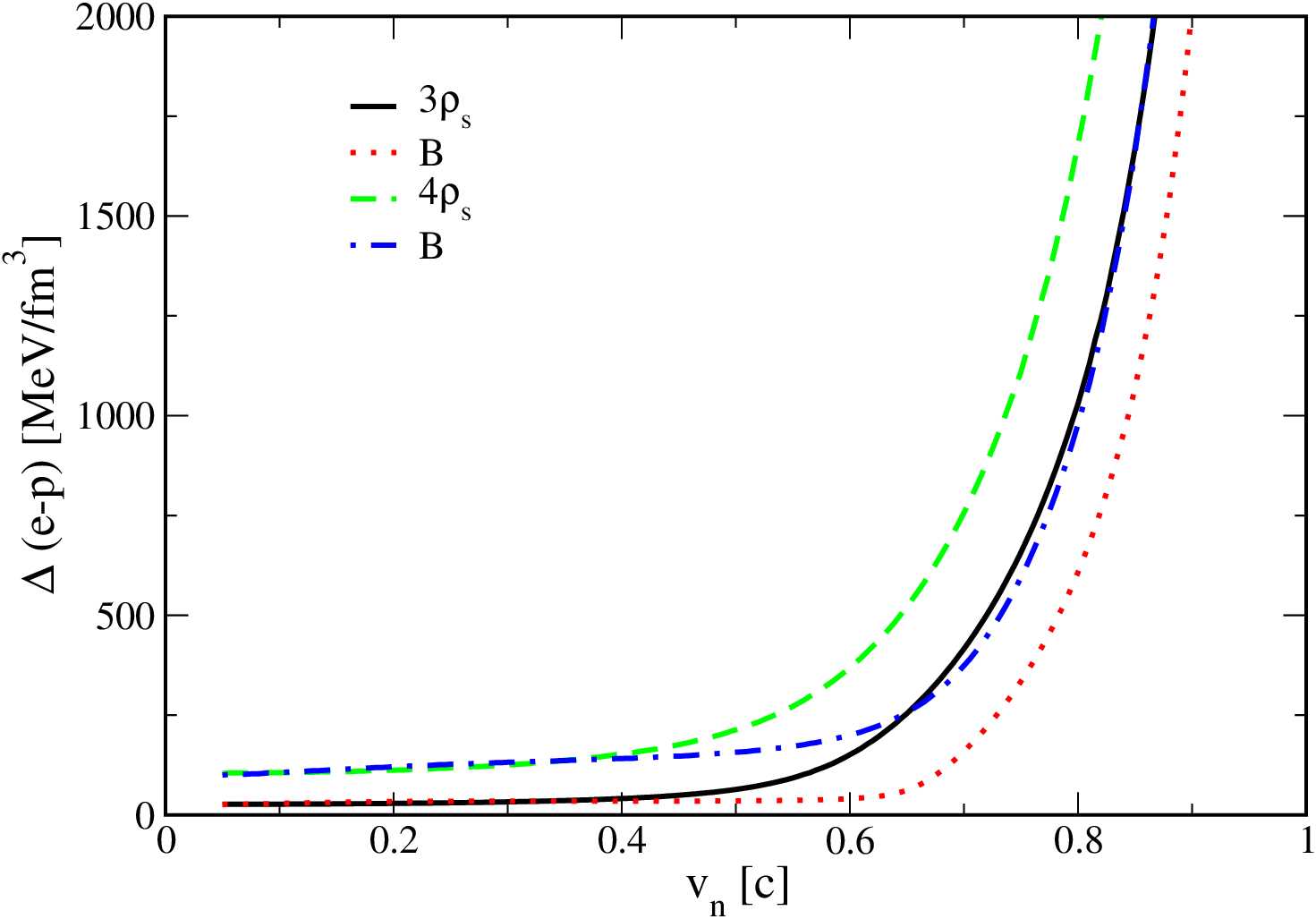}}
\caption{(Color online) a) The incoming matter velocity $v_n$ and the front velocity $v_f$ in the global frame (QM rest frame) are shown as a function of $\rho_b$. The solid and the 
dashed lines are for $v_n$ without and with magnetic effect and the dotted and dash-dotted lines are for corresponding $v_f$'s. Curves are plotted for $v_h=0.3$. \
b) The difference of $(e-p)$ for QP and NP is plotted as a function of $v_n$. The solid line (without magnetic effect) and the dotted line
(with magnetic effect) are plotted
for $\rho_b=3\rho_s$ and the dashed and dash-dotted curves are for $\rho_b=4\rho_s$.}
\label{vel2}
\end{figure}

The global nuclear matter speed and the front velocity is an important parameter for this PT.
The value of $v_n$ depends both on $v_h$ and $v_q$. Although $v_h$ is kept constant for our calculation $v_n$ changes with $\rho_b$ as the value of $v_q$ changes. Therefore, 
it is interesting to see how the front velocity $v_f$ and $v_n$ changes with $\rho_b$. In fig \ref{vel2}a we have plotted the $v_n$ and $v_f$ as a function of $\rho_b$ for $v_h=0.1$. 
At low densities 
$v_n$ first decreases and attains a minimum negative value at $2\rho_s$. Beyond this point $v_n$ gradually increases and again becomes positive beyond $2.5 \times \rho_s$ and 
then gradually attains a constant value at higher densities. The nature of 
$v_f$ is completely opposite, it first increases at low densities and attains a maximum value at $2\rho_b$ ($v_f=0.1=v_h$) and then gradually decreases to attain almost a 
constant value at 
higher densities (which is smaller than $v_h$). In the region of our interest ($\rho_b > 3\rho_s$) $v_f$ is always greater than $v_n$ and they are both smaller than $v_h$.
The nature of both $v_n$ and $v_f$ are the same for a magnetic star only their corresponding values are smaller than the non-magnetic star. 

Next in fig \ref{vel2}b we check the variation of $\Delta (e-p)$ with $v_n$ for two values of $\rho_b$. The value is always positive implying that the burning process at 
such densities is always a deflagration. As the velocity increases the $\Delta (e-p)$ increases, implying a strong deflagration. 
$\Delta(e-p)$ is always greater for normal NS compared to magnetars implying that for magnetars deflagration is slow.

\subsection{Taub-adiabat}

\begin{figure}
\subfloat[]{\includegraphics[width = 2.4in]{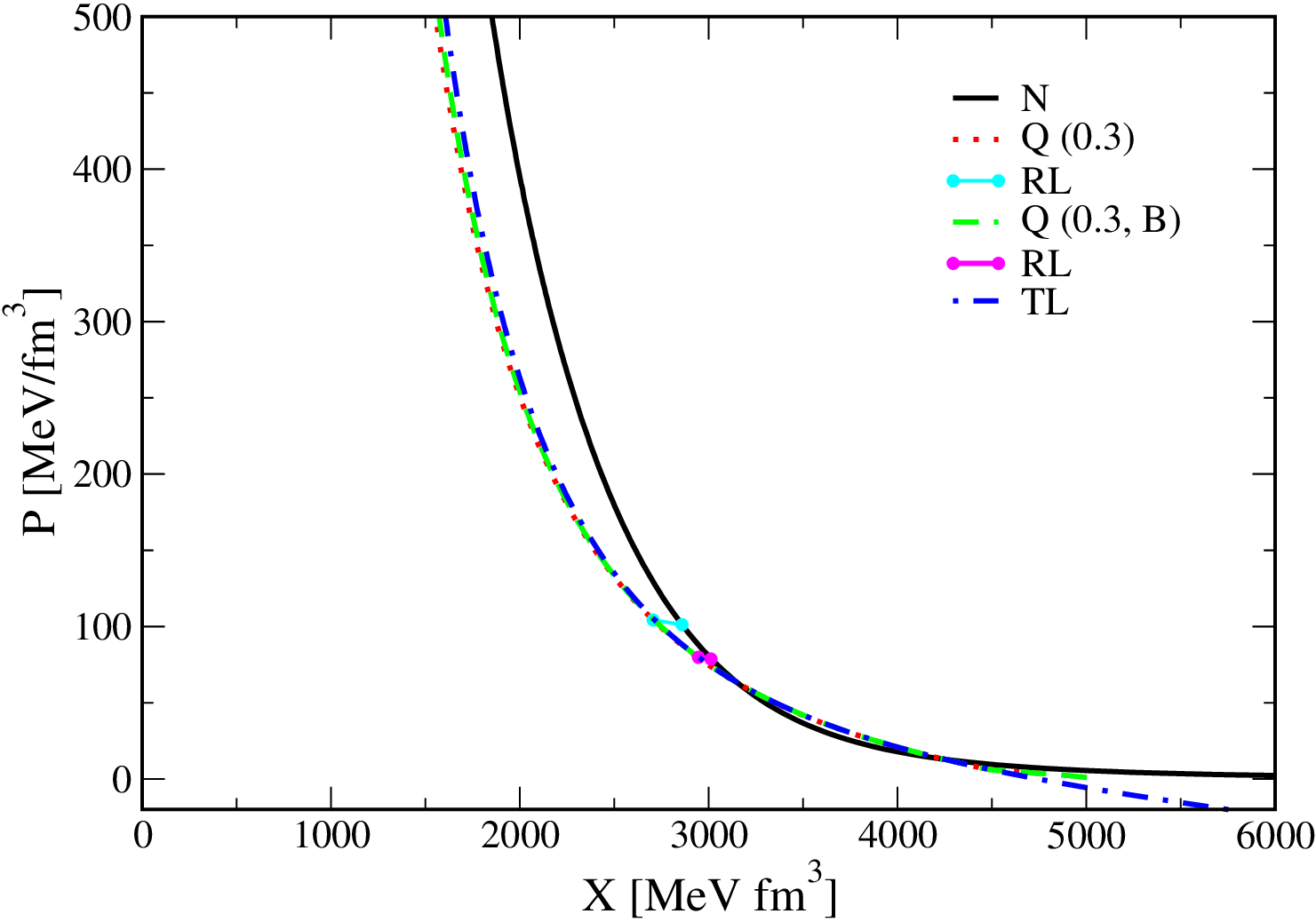}} \quad 
\subfloat[]{\includegraphics[width = 2.4in]{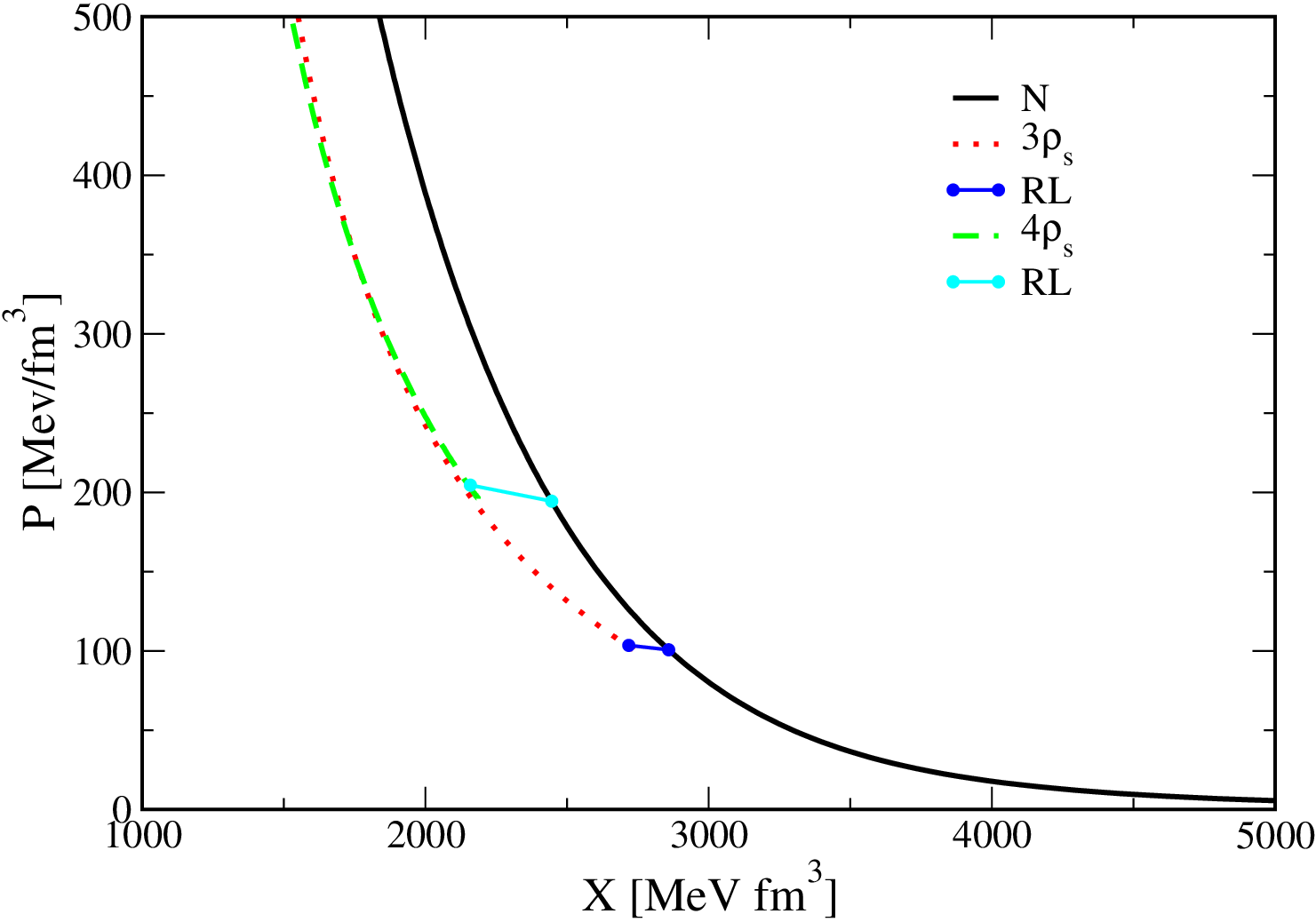}}
\caption{(Color online) TA are shown in the $X-P$ plane. a) Curves are plotted for a fixed $v_h=0.3$ and by varying the density.
The horizontal lines connected with dots are the so-called RL, showing PT from NM to QM for $\rho_b=3\rho_s$. TA and RL with and without magnetic field induced EoS are shown.
The RL for magnetic field induced EoS happens at much lower pressure.
Also shown in the graph the TA for tL shock adiabat (blue dash-dot line).
b). TA obtained by varying $v_h$ for two fixed $\rho_b$, $3\rho_s$ (dotted) and $4\rho_s$ (dashed). 
The slant horizontal lines are for RL for $v_h=0.1$.}
\label{taub2}
\end{figure}

Nuclear to quark PT can also be realized if we plot the Taub adiabat (TA) curves. TA is a single equation which can be obtained from the conservation equations and reads as
\begin{equation}
 (p_{n}+\varepsilon_q) X_q=(p_q+\varepsilon_n) X_n\,,
 \label{eqta}
\end{equation}
where $X_i=w_i/n_i^2$. The thermodynamic quantities of a given phase can be regarded as a function of this $X$. For a given initial state of NM
(a fixed point in the curve) one can have a TA of the QM by a line in the \mbox{$X_q-p_q$} plane. The slope of the RL, 
connecting this initial point in the NM with the point $(X_q,p_q)$ on the TA is related to the incoming velocity $v_h$. 
As $v_h$ increases the slope of the line increases as $(\gamma_h v_h)^2$ \cite{landau}. Therefore, for each $v_h$
there is a specific point on the TA corresponding to the state of compressed QP. The discontinuity across the PT front can be understood from the TA. The RL represents
the pressure, energy density and density discontinuity across the front from NM to QM.

In fig \ref{taub2}a we have plotted such a TA. The black line represents the initial 
NM EoS plotted in this plane. The shock adiabat (red dotted curve) is obtained by varying the nuclear density for a fixed incoming matter velocity ($v_h=0.3$) 
using the conservation conditions.
Smaller density implies higher X, and as density increases, X becomes smaller. We find that at lower densities the shock adiabat can be on the left side of the nuclear curve not observed for equilibrium PT. As the density increases, it goes to the right of the nuclear adiabat and their difference increases. The blue line represents the 
RL. 
As the HM EoS is stiffer at higher energies than QM EoS therefore usually it lies on the right of the shocked matter. However, 
there is a small window where it rests on the left of the shock adiabat (at low energies the curves cross in both fig \ref{eos}a and fig \ref{taub2}a)). The shock adiabat can be identified with non-horizontal RL where both density and pressure changes. 

The magnetic shock adiabat shows a similar behavior, and at lower densities, it almost overlaps with the non-magnetic curve. The curve differs at higher densities.
It is interesting to note that the density range where the shock adiabat lies on the left of the nuclear curve coincides with the density range where the burning becomes detonation. 
This is a fascinating phenomena which shows up in almost all the figures. The 
RL for the magnetic adiabat lies at a lower pressure than that for non-magnetic adiabat which is expected as the magnetic EoS is softer than the non-magnetic one. 

In fig \ref{taub2}b the shock adiabat is obtained by varying $v_h$ for fixed values of $\rho_b$. The RL are shown for $v_h=0.3$. Curves have been plotted for two densities 
$3\rho_s$ and $4\rho_s$. The curve for smaller density starts from lower pressure as expected. Therefore, the RL slope is also softer than that for $4\rho_s$. 
The magnetic curve shows a similar feature and has not been shown in the figure as it does not add any new physics to it.

\section{Summary and Conclusion}
In this present article, we mainly focus our attention on the effect of magnetic field on PT of an NS to QS. We have used relativistic MHS 
conservation condition to study this effect along with magnetic field induced EoS. For simplicity, we have chosen the HT frame (where the magnetic and matter 
velocities are aligned in the rest frame of the front)
and also assumed infinite conductivity. We aim to categorize the PT process, whether it is a fast burning detonation or a slow deflagration. 
In our calculation, we have used the hadronic and quark EoSs which is
consistent with the recent constraints. We have assumed magnetic fields of strength $10^{15}$ G at the surface of the star which is usually associated with magnetars. The central magnetic 
field is considered to be of the order of 
$1.7 \times 10^{18}$ G. The energy of NM is higher than that of QM (at fixed number density) at higher densities, and the process can be exothermic.
Therefore, the PT induced by a shock like discontinuity at the star center will propagate outwards, converting NM to QM. 

Observing the criterion for detonation and deflagration, we found that the velocity of the NM in the majority of the density range is higher than that of the QM, 
which is the condition for a deflagration. Also, the $\Delta(e-p)$ comparison of the respective phases comes to the same conclusion. 
The burning process at the star center most likely starts as a deflagration process. However, in almost all the curves we have found that there is a small density window at 
lower densities where the process can be a detonation one. In this little window, all the parameters behave differently. 
The criterion for detonation and deflagration depends strongly on our choice of EoS and to infer further, studies employing different EoSs should be carried out. 

Most of the exciting and new physical insight comes when we compared the magnetic field of the unburnt NM and burnt QM. At small enough infalling matter velocities the 
resultant magnetic field of the QS is lower than that of the NS. However, for higher values of infalling matter
velocities, the magnetic field of QM becomes larger. Therefore, depending on the initial density fluctuation and on whether the PT is a violent one or not the QS could
be more magnetic or less magnetic. This can have substantial observational significance because a strong magnetar can suddenly become less magnetic and will
not show common magnetar properties like anomalous x-ray pulses and flares. On the other hand, a regular NS can suddenly start to exhibit x-ray pulses and giant flashes and other 
magnetar characteristics, and this change happens suddenly as the PT is a fast process.

The sudden PT can also have a massive effect on the magnetic tilt of the star. For smaller velocities and densities the magnetic inclination are not affected much but for higher
infalling velocities the tilt of the MA can change suddenly with PT. In such extreme cases, the magnetic angle suddenly flips sign and even can increases a 
lot suddenly. All previous calculations regarding the evolution of the magnetic tilt axis is a slow process and is connected with its lifetime and speed. However, the change 
in the magnetic tilt for magnetars due to PT is a sudden process and is an interesting one,
implying that a star with MA tilted to the right can undergo a PT and the resultant QS can have an MA leaned to the left?
This can have considerable effects on the observation of pulsars. A pulsar previously recorded can undergo a PT and can completely disappear.
On the other hand, we can suddenly identify a new pulsar in the sky after it has undergone PT without any supernovae happening in the near past.

Although we have in detail discussed the magneto-hydrostatic scenario of the PT of NS to QS, we have only established the initial conditions and the possible PT mechanism.
The actual dynamic of the PT would be complete once we study and understand the magneto-hydrodynamic PT scenario solving the dynamic Euler's equations. Although such calculation 
might be very involved, it is on our immediate agenda \cite{mallick-dynamical}.

\section{Authors contributions}
RM would like to thank SERB, Govt. of India for monetary support in the form of Ramanujan Fellowship (SB/S2/RJN-061/2015) and Early Career Research Award (ECR/2016/000161). 
RM and AS would like to thank IISER Bhopal for 
providing all the research and infrastructure facilities.

\end{document}